\documentclass[numbook,envcountsect,envcountsame,envcountreset,
 smallextended
]{svjour3} \smartqed
\usepackage{graphicx}
\usepackage{mathptmx}
\usepackage{amssymb}
\usepackage{amsmath}
\usepackage{graphicx,epsf}
\usepackage{epstopdf}

\spnewtheorem{chart}[theorem]{Chart}{\bf}{\rm}

\spnewtheorem{lexample}[theorem]{Example}{\bf}{\rm}
\spnewtheorem{lremark}[theorem]{Remark}{\bf}{\it}

\renewcommand{\figurename}{Diagram}
\renewcommand{\tablename}{Chart}
\renewcommand{\captionstyle}{\normalsize}

\newcommand{\fVPS}{(f\kern1pt VPS)}

\journalname{ArXiv.org} 

\begin{document}
    \raggedbottom
    \allowdisplaybreaks
\title{\bf\it Three-Dimensional Stationary Spherically Symmetric Stellar Dynamic
 Models Depending on the Local Energy
}

\titlerunning{Three-Dimensional Stationary Spherically Symmetric Stellar Dynamic}       

\author{J\"{u}rgen Batt   \and  Enno J\"{o}rn   \and    Alexander\,L.~Skubachevskii}


\institute{J\"urgen Batt and Enno J\"orn, Mathematisches Institut der Universit\"at M\"unchen,\\ 
80333 M\"unchen, Theresienstr.39, Germany 
\qquad \email{batt@mathematik.uni-muenchen.de}
                    \\
                Alexander\,L.~Skubachevskii,
                            Peoples Friendship University of Russia (RUDN
                                        University),\\
              6~Miklukho-Maklaya Street, Moscow, 117198, Russian Federation
\qquad              \email{skub@lector.ru}           
%
}

\date{Received: {\bf 04.07.2021} / Accepted: date}  

\maketitle

\noindent {\bf Abstract} \  \\  [2ex]
    %
The stellar dynamic models considered here are triples ($f,\rho,U$) of three
functions: the distribution function $f=f(r,u)$, the local density $\rho=\rho(r)$ and
the Newtonian potential $U=U(r)$, where $r:=|x|$, $u:=|v|$
\big($(x,v)\in\mathbb{R}^3\times\mathbb{R}^3$ are the space-velocity coordinates\big),
and $f$ is a function $q$  of the local energy $E=U(r)+\dfrac{u^2}2$. Our first result is
an answer to the following question: Given a (positive) function $p=p(r)$ on a bounded
interval $[0,R]$, how can one recognize $p$ as the local density of a stellar dynamic
model of the given type (``\,inverse problem'')? If this is the case, we say that $p$ is
``extendable'' (to a complete stellar dynamic model). Assuming that $p$ is strictly
decreasing we reveal the connection between $p$ and $F$, which appears in the nonlinear
integral equation $p=FU[p]$ and the solvability of Eddington's equation between $F$ and
$q$ (Lemma \ref{t4.1}). Second, we investigate the following question (``direct
problem''): Which $q$ induce distribution functions $f$ of the form $f=q(-E(r,u)-E_0)$ of
a stellar dynamic model? This leads to the investigation of the nonlinear equation
$p=FU[p]$ in an approximative and constructive way by mainly numerical methods.
 --- The paper extends preceding work on flat galaxies \cite{1} to the three-dimensional case.
In particular, the present answer to the extendability problem is completely different as in \cite{1}. 
The present paper also opens the way to further explicit solutions of the 
Vlasov-Poisson system beyond the classical known examples which are
for instant given in  \cite{3} .
\newpage
\tableofcontents


    \section{Introduction}\label{intro}

The Vlasov--Poisson System (VPS) in 3 dimensions (stellar dynamic version) has the
following form:
    \begin{align}
\dfrac{\partial{f}}{\partial{t}}+v\cdot\dfrac{\partial{f}}{\partial{x}}
-\dfrac{\partial}{\partial{x}}\, U(t,x)\cdot\frac{\partial{f}}{\partial{v}}&=0,\tag{V}
    \\
\Delta U(t,x)&=4\pi\rho(t,x) \tag{$P_1$}
    \\
\text{or} \quad U(t,x)&=-\int\dfrac{\rho(t,y)}{|x-y|}\,dy, \tag{$P_2$}
    \\
\rho(t,x)&=\int f(t,x,v)\,dv.\tag{D}
    \end{align}
Here $f=f(t,x,v)\ge 0$ is the distribution function of the gravitating matter,\\
$U=U(t,x)\le0$  the Newtonian potential and $\rho(t,x)\ge0$  the local density. The
system has been intensively investigated in many directions. For the case of
time-dependent functions (initial value problem), \cite{8} gives a survey until 2007. The
stationary spherically symmetric functions are characterized by the property
$f(x,v)=f(A_1x,A_2v)$ for all $A_1,A_2\in S0(3)$; for a short account of this class,
relevant for our work, see \cite{1}, also for references.

The aim of the present paper is twofold. Our first problem is known as the ``inverse
problem'': to identify those functions $p$, defined on a bounded interval $[0,R]$, as the
local density of a stationary spherically symmetric stellar dynamic model, in which $f$
depends on the local energy:
    \[
f(r,u)=q(-E-E_0),  \qquad  \text{where} \quad  E_0>0 \quad \text{is a constant}.
    \]
This question occurs if one wants to determine the three quantities $f$, $\rho$, $U$ from
observation. The result of an observation generally is a brightness profile, which, by
certain strategies, can be turned into a mass profile. The question which follows is the
determination of the potential $U$ and the distribution $f$ (Sections
\ref{s2}--\ref{s5}).

Our second problem is called the direct problem. It is known that the distribution $f$ of
a stationary spherically symmetric stellar dynamic model is a function of the local
energy $E$ and the angular momentum $F:=x^2v^2-(xv)^2$ (this fact is called Jeans'
theorem) \cite{2}. The direct problem partially poses the opposite question, namely:
which functions $q$ admit finding functions $\rho(r)$ and $U(r)$ together with a constant
$E_0>0$ such that $f(r,u)=q(-E-E_0)$, $\rho$ and $U$ form a triple of a stationary
spherically symmetric stellar dynamic model (Sections~\ref{s6}--\ref{s8}). We give a
short overview over the different sections.

Section \ref{s2}: Introduction of the potential operator $U=Lp$ on its domain
$\mathcal{D}(L)$ (Definition~\ref{d2.1}) with its elementary properties (Lemma \ref{l2.1}). 
Each strictly decreasing function $p\in\mathcal{D}(L)$ satisfies a nonlinear
integral equation $p=FLp$ with an appropriate $F=F[p]$ (Lemma \ref{l2.2}).

Section \ref{s3}: Definition of the stationary spherically symmetric solutions depending
on the local energy and proofs of their properties, 
Equivalence Lemma and Eddington's equation (Lemma \ref{l2.1}). 

Section \ref{s4}: The inverse problem: its formulation and its solution
(Theorem~\ref{t4.2}).

Section \ref{s5}: Presentation of examples,  which illustrate Theorem \ref{t4.2}, and the
concept of extendability.

Section \ref{s6}: Formulation of the direct problem and its conversion into the
equivalent problem of solving a nonlinear integral equation of the form
    \[
Lp-E_0=G_0(p).
    \]

Section \ref{s7}: Construction of an approximating nonlinear system (ANS) of the form
    \[
\sum_{k=0}^{n-1} A_{ik}x_k:=\sum_{k=0}^{n-1} B_{ik}x_{k}-C_kx_k=G_0(x_i)
    \]
and calculation of the matrix ($A_{ik}$).

Section \ref{s8}: Numerical analysis of the (ANS), description of the approximation and
convergence, examples.

Section \ref{s9} (Appendix): Contains Tonelli's work on Abel's and Eddington's equations
with full proofs.

Section \ref{s10}: Contains suggestions for further research.


\section{The potential operator in spherical symmetry}\label{s2}

We define the potential operator
    \[
Lp(x):=\int_{\mathbb{R}^3}\frac{p(y)}{|x-y|}\,dy, \qquad x\in\mathbb{R}^3,
    \]
for certain functions $p$ on $\mathbb{R}^3$, which are spherically symmetric. This means
(by abuse of notation) that $p(x)=p(r)$, $r:=|x|$. We first conclude that $Lp$ is also
spherically symmetric . In fact, if $A\in SO(3)$, then, assuming that $y=Az$, we have
    \begin{align*}
Lp(Ax)&=\int_{\mathbb{R}^3}\frac{p(y)}{|Ax-y|}\,dy=\int_{\mathbb{R}^3}
\frac{p(y)}{\big|A(x-A^{-1}y)\big|}\,dy= \int_{\mathbb{R}^3}\frac{p(Az)}{|x-z|}\,dz
    \\
&=\int_{\mathbb{R}^3}\frac{p(z)}{|x-z|}\,dz =Lp(x).
    \end{align*}

 We define\;\; $\mathbb{R}_{0+}:=\big\{r\in\mathbb{R}\colon$ 
$r\ge0\big\}$\;\; and\;\; $\mathbb{R}_{+}:=\big\{r\in\mathbb{R}\colon r>0\big\}$.
    \begin{definition}\label{d2.1}
Let $\mathcal{D}(L)$ be the set of functions
$p\colon\mathbb{R}_{0+}\to\mathbb{R}_{0+}\cup\{\infty\}$ with the following properties:
    \begin{itemize}
\item[\rm (a)] $p\in C(\mathbb{R}_+)$,
     \\
\item[\rm (b)] for all $r>0$ we have $\displaystyle\int_0^r p(s)s^2\,ds<\infty$,
$\displaystyle\int_r^\infty p(s)s\,ds<\infty$,
    \\
\item[\rm (c)] there exists a $\delta>0$ such that
 for all $r\in (0,\delta)$, we have $p(r)>0$.
    \end{itemize}
    \end{definition}
    \begin{lemma}\label{l2.1}
For $p\in\mathcal{D}(L)$, we have:
    \begin{align}
& 1) \ Lp(r)=4\pi\left[\frac1r\int_0^rp(s)s^2\,ds+\int_r^\infty p(s)s\,ds\right], \qquad
r>0, \label{e2.1}
    \\
& 2) \ Lp\in C^2(\mathbb{R}_+), \quad \text{and} \notag
    \\
&\phantom{2)}\ (Lp)'(r)=-\frac{4\pi}{r^2}\!\int_0^r p(s)s^2\,ds, \quad r>0,\label{e2.2}
    \\
&\phantom{2)}\ (Lp)''(r)=-\frac 2r(Lp)'(r)-4\pi p(r), \qquad r>0. \label{e2.3}
    \\
& 3) \ Lp>0 \ \text{and} \ (Lp)'<0, \ \text{that is}, Lp \ \text{is strictly decreasing
on} \ \mathbb{R}_+. \ \text{Because }\notag
    \\
&\phantom{2)}\ \text{the limits}\notag
    \end{align}
    \[
Lp(0):=\lim_{r\to0} Lp(r), \qquad Lp\,(\infty)=\lim_{r\to\infty} Lp(r)
    \]
exist, the function $Lp$ has a strictly decreasing inverse
    \[
(Lp)^{-1}\colon \big(Lp(\infty), Lp(0)\big)\to (0,\infty).
    \]
    \end{lemma}
    \textbf{Proof.}
1)~Using spherical coordinates
    \[
x=(r\sin\psi\cos\varphi, r\sin\psi\sin\varphi, r\cos\psi),
    \]
we have
    \begin{multline}
Lp(x)=\int_0^\pi\!\int_0^{2\pi}\!\int_0^\infty\frac{p(s)\sin\psi}
{\sqrt{r^2+s^2-2rs\cdot\cos\psi}}\,s^2\,ds\,d\varphi\,d\psi
    \\
=2\pi\int_0^\infty\,\int_0^\pi\frac{\sin\psi}{\sqrt{r^2+s^2-2rs\cdot\cos\psi}}\, d\psi
p(s)s^2\,ds.\notag
    \end{multline}
In the inner integral we substitute $u:=\sqrt{r^2+s^2-2rs\cos\psi}$ and get
    \begin{align*}
\int_0^\pi\frac{\sin\psi}{\sqrt{r^2+s^2-2rs\cdot\cos\psi}}\,d\psi&=
\!\!\!\int_{\sqrt{r^2+s^2-2rs}}^{\sqrt{r^2+s^2+2rs}}\!\!\!1\cdot du\cdot\frac1{rs}
   \\
&=\frac{(r+s)-|r-s|}{rs}=\begin{cases}
                        \dfrac{2}{r}\quad &\text{for}\quad s\le r,
                            \\[0,7em]
                        \dfrac 2s \quad &\text{for}\quad s\ge r,
                        \end{cases}
    \end{align*}
and \eqref{e2.1}  follows.

2)~\eqref{e2.2} results from differentiating \eqref{e2.1}, and \eqref{e2.3} follows from
differentiating~\eqref{e2.2}.

3)~Inequality $Lp>0$ is a consequence of Definition~\ref{d2.1} c), and $(Lp)'<0$ results
from \eqref{e2.2}. The existence of the limits and of the inverse operator are direct
consequences of these facts.\hfill $\square$
    %
\\[1ex]
Most of our functions $p\in\mathcal{D}(L)$ will have compact support.\\[1ex]
 We define\\[-4ex]
\begin{center} 
$\mathcal{D}_R(L):= \{p\in\mathcal{D}(L)\colon p>0$ on $[0,R)$, $p=0$ on $[R, \infty)$\}.\\
$\mathcal{D}_R^-(L):=\{p\in\mathcal{D}_R(L)\colon p$ strictly decreasing on$[0,R)\}$.
\end{center} 
The functions $p\in\mathcal{D}_R^-(L)$ are solutions of a nonlinear integral equation, as
the following lemma shows.
%
 \begin{figure}[ht]\label{D21}
        \begin{minipage}{100mm}\caption{  for\; $F(h)=p\circ(Lp)^{-1}(h)$, \;($E_0:=Lp(R)$).  }
    \centering\includegraphics[scale=0.59]{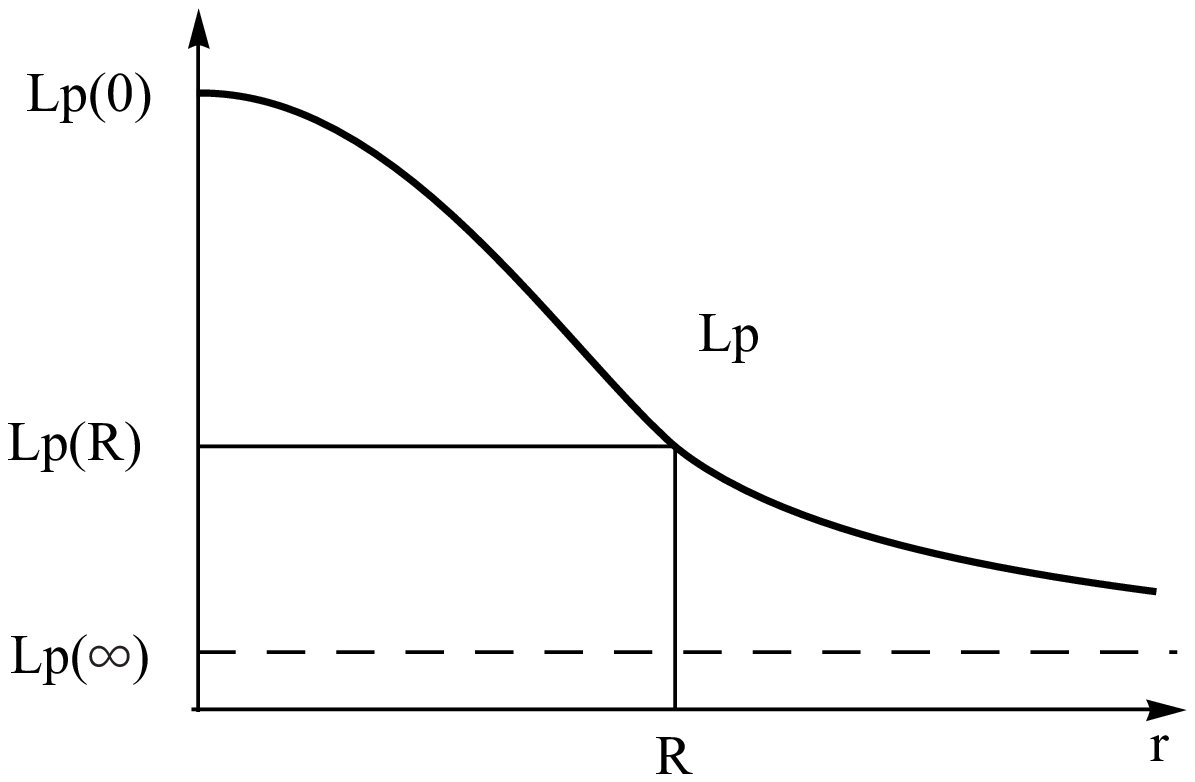}
   \end{minipage}
    \\[3mm]
    \begin{minipage}{100mm}
    \centering\includegraphics[scale=0.59]{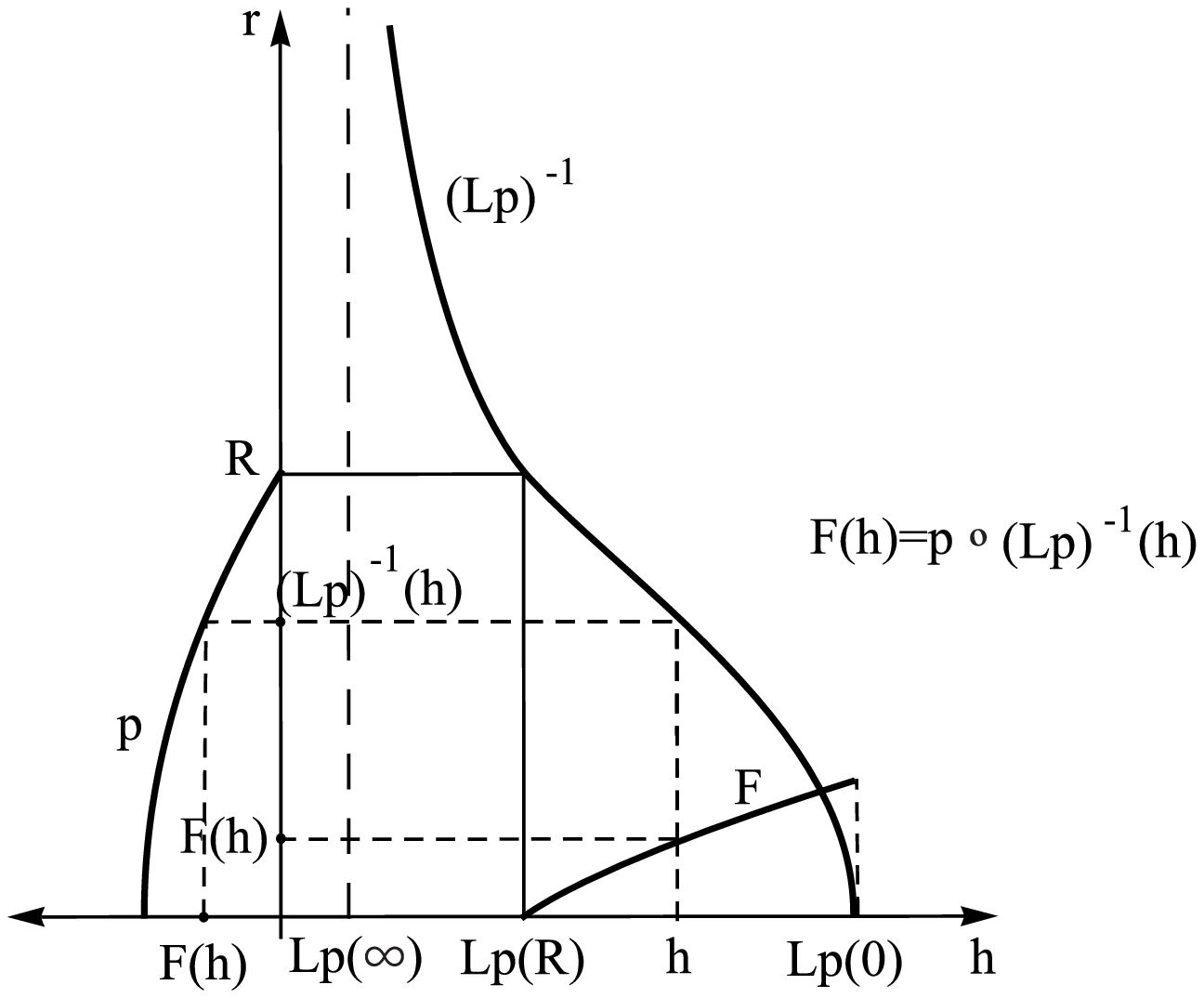}
    \end{minipage}
     \end{figure}
$~$\\[2ex]
    \begin{lemma}\label{l2.2}
Let $p\in\mathcal{D}_R^-(L)$.
\\
Then there exists a unique strictly increasing function
   \[
F:=F[p]\colon \big[Lp(R), Lp(0)\big)\to p\big((0,R]\big)
    \]
such that
    \[
p(r)=F\circ Lp(r), \qquad r\in(0,R].
    \]
    \end{lemma}
    \textbf{Proof.}
Lemma \ref{l2.1}\, 3) says that
    \[
Lp\colon (0,R]\to \big[Lp(R), Lp(0)\big)
    \]
is strictly decreasing and has a strictly decreasing inverse
    \[
(Lp)^{-1}\colon \big[Lp(R), Lp(0)\big)\to(0,R].
    \]
Because $p$ is strictly decreasing on $(0,R]$, the composition
    \[
F:=p\circ(Lp)^{-1}\colon \big[Lp(R), Lp(0)\big) \to p((0,R])
    \]
exists and is strictly increasing (see  Diagram~2.1) 
Then
    \[
F(h)=p\circ(Lp)^{-1}(h),\qquad h\in[Lp(R),Lp(0))
    \]
implies
    \[
F\circ Lp(r)=p(r), \qquad r\in(0,R].
    \]
The uniqueness of $F$ is immediate: 
If $G$ satisfies $p=G\circ Lp$, then $G=p\circ(Lp)^{-1}=F$. \hfill$\square$
    \begin{corollary}\label{co2.1}
Let $F(h):=0$ for $h\in(0,Lp(R))$. Then under the conditions of Lemma {\rm\ref{l2.2}} we
have
    \[
p(r)=F\circ Lp(r) \ \text{on} \ (0,R]\longleftrightarrow p(r)=F\circ Lp(r) \ \text{on} \
\mathbb{R}_{+}.
    \]
    \end{corollary}
    \textbf{Proof.}
    Let $p(r)=F\circ Lp(r)$ on $(0,R]$. In this case, if $R<r$, then $p(r)=0$, and $Lp(R)>Lp(r)$
implies that $F\circ Lp(r)=0$, i.\,e. $p(r)=F\circ Lp(r)$ on $\mathbb{R}_+$. The inverse
statement is trivial.\hfill$\square$
\\ Often we use the abbreviation $P:=Lp$.


    \section{Stationary spherically symmetric solutions depending
on the local energy and their properties}\label{s3}
    \begin{definition}\label{d3.1}
A triple $(f,\rho,U)$ of functions $f=f(r,u)$, $\rho=\rho(r)$, $U=U(r)$ is called a
stationary spherically symmetric $E$-dependent solution of the (VPS) if
$\rho\in\mathcal{D}_R^-(L)$ and there exists a function $q=q(s)$ with the following
property:
    \begin{align*}
&(Q)\quad q\in L_{\rm loc}^1(\mathbb{R}), \;\;  q(s)=0 \;\; \text{for} \
s\in(-\infty,0], \quad q(s)>0 \;\; \text{for} \;  s\in(0,P(0)\!-\!E_0)
    \\
    & \text{such that}
    \\
&(V)\quad f(r,u)=q\big(-E(r,u)-E_0\big),
    \\
&\qquad \;\; E(r,u):=U(r)+\frac{u^2}2, \ u:=|v|,
    \\
&\qquad \;\; (\text{we write} \ f=f_q),
    \\
&(P)\quad U(r)=-L\rho(r), \ L\rho(R)=:E_0,
    \\
&(D)\quad \rho(r)=\int_{\mathbb{R}^3}f(r,|v|)\, dv, \qquad r\in\mathbb{R}_{+}.
    \end{align*}
    \end{definition}

We note that $(Q)$ states the properties of $q$, $(V)$ refers to $f\,'s$ being an
integral of Vlasov's equation (i.e., being constant along the characteristics), $(P)$ is
the integrated form $(P_2$) of Poisson's equation, $(D)$ is the definition of the local
density.

As a preparation for the following important lemma we prove a crucial identity.
    \begin{lemma}\label{l3.1}
Let $p\in\mathcal{D}(L)${\rm,} $E_0>0${\rm}\; and $q$ satisfy $(Q)$. Then for $E_0\le
h<Lp(0)$ the following equality holds{\rm:}
    \begin{equation}\label{e3.1}
\int_{\mathbb{R}^3}\!q\left(h-E_0-\frac{v^2}2\right)dv= 4\pi\sqrt{2}\int_0^{h-E_0}
\!\!\!q(s)\sqrt{h-E_0-s}\,ds.
    \end{equation}
    \end{lemma}
    \textbf{Proof.}
Since $q(s)=0$ for $s\in(-\infty,0]$, we have
    \begin{align*}
\int_{\mathbb{R}^3}\!\!q\left(h-E_0-\frac{v^2}2\right)dv&= \int_{|v|<\sqrt{2(h-E_0)}}\;
q\left(h-E_0-\frac{v^2}2\right)dv
    \\
&=\int_0^\pi\!\!\int_0^{2\pi}\!\!\int_0^{\sqrt{2(h-E_0)}}\!\!\!q\left(h-E_0\!-\!
\frac{u^2}2\right) u^2\,du\, d\varphi\sin\psi\,d\psi
    \\
&=4\pi\int_0^{\sqrt{2(h-E_0)}}\!\!q\left(h-E_0-\frac{u^2}2\right)u^2\,du.
    \end{align*}
Passing to the new variable $s:=h-E_0-\dfrac{u^2}2$, we have $u=\sqrt{2(h-E_0-s)}$\,.
Hence we get \eqref{e3.1}.\hfill$\square$
    \begin{lemma}\label{l3.2} {\bf(\rm\textbf{Equivalence Lemma})}
    \\[1ex]
{\rm (a)}~Let $(f_q,\rho, U)$ be a stationary spherically symmetric $E$-depending
solution of the (VPS). Let $F:=F[p\,]$ {\rm(}Lemma {\rm\ref{l2.2}),} where $p=\rho$. Then
    \begin{equation}\label{e3.2}
F(h)=4\pi\sqrt{2}\int_0^{h-E_0}\!q(s)\sqrt{h-E_0-s}\,ds \quad \text{for} \quad
h\in\big[E_0, P(0)\big).
    \end{equation}
{\rm (b)}~Let $q$ satisfy $(Q)${\rm,} and let
    \begin{equation}\label{e3.3}
F(h):=4\pi\sqrt2\int_0^{h-E_0}\!\!q(s)\sqrt{h-E_0-s}\,ds \quad \text{for} \quad
h\in\big[E_0, P(0)\big).
    \end{equation}
Assume the integral equation
    \begin{equation}\label{e3.4}
p(r)=F\circ Lp(r)
    \end{equation}
has a solution $p\in\mathcal{D}_R^-(L)$ on $\mathbb{R}_{+}$. We define $\rho:=p${\rm,}
$U(r):=-L\rho(r)${\rm,} $E_0:=Lp(R)${\rm,} and $f_q(r,u):=q(-E(r,u)-E_0)$. Then $(f_q,
\rho,u)$ is a stationary spherically symmetric $E$-depending solution of the
{\rm(}VPS{\rm)}.
    \end{lemma}
    \textbf{Proof.}
$~$\\[-2ex]

a)~Let $(f_q,\rho, U)$ be a stationary spherically symmetric $E$-depending solution.
Then, by virtue of Lemma \ref{l2.2}, $(D)$, $(V)$, and \eqref{e3.1}, we have
    \begin{align*}
F\circ L\rho(r)=\rho(r)&=\int_{\mathbb{R}^3}\!f_q\big(r, |v|\big)\,dv=
\int_{\mathbb{R}^3}\!q\left(L\rho(r)-E_0-\frac{v^2}2\right)dv
    \\
&= 4\pi\sqrt2\int_0^{L\rho(r)-E_0}\!\!q(s)\sqrt{L\rho(r)-E_0-s}\,ds \quad \text{for}
\quad r\in(0,R],
    \end{align*}
and \eqref{e3.2} follows.

b)~Our assumptions imply that $(Q)$, $(P)$, $(V)$ are satisfied. Furthermore, by virtue
of \eqref{e3.4}, \eqref{e3.3}, \eqref{e3.1}, and $(V)$, we have
    \begin{align*}
\rho(r)=p(r)=F\circ Lp(r)&=4\pi\sqrt2\int_0^{Lp(r)-E_0}\!\!q(s) \sqrt{Lp(r)-E_0-s}\, ds
    \\
&= \int_{\mathbb{R}^3} q\left(Lp(r)-E_0-\frac{v^2}2\right)dv
    \\
&=\int_{\mathbb{R}^3} \!f(r,u)\, dv, \qquad r\in(0,R].
    \end{align*}
Hence $(D)$ is also satisfied.\hfill$\square$
\\[-1ex]

In the sequel, we will use the definition
    \[
F_0(h):=F(h+E_0) \quad \text{on} \quad [\,0,P(0)-E_0).
    \]
Then \eqref{e3.2} has the form
    \begin{equation}\label{e3.5}
F_0(h)=4\pi\sqrt2\int_0^h q(s)\sqrt{h-s} \, ds, \quad h\in[0, P(0)-E_0).
    \end{equation}
This is an equation of the form
    \[
g(x)=\int_0^x\!\!f(s)\sqrt{x-s} \, ds,
    \]
which is called Eddington's equation. The results on its solvability are based on the
theory for an equation of the form
    \[
g(x)=\int_0^x\!\!\frac{f(s)}{\sqrt{x-s}} \, ds,
    \]
which is called Abel's equation. It was Tonelli \cite{11}, who has given existence proofs
for these equations (a review of his work is given in \cite{5}).

For the sake of the completeness of the present work, the main results and their proofs
are given in the Appendix.


    \section{The inverse problem}\label{s4}
In this section we consider and solve the following question: Given a function
$p\in\mathcal{D}_R^-(L)$, under which conditions is $p$ the local density of a stationary
spherically symmetric $E$-dependent solution? In this case we say ``$p$ is extendable''
(by $f$ and $U$ to a stationary  spherically symmetric $E$-dependent solution).

The following proposition gives a first necessary and sufficient criterion that a given
$p\in\mathcal{D}_R^-(L)$ is extendable.
    \begin{theorem}\label{t4.1}
Let $p\in\mathcal{D}_R^-(L)$. Then $p$ is extendable if and only if Eddington's equation
\eqref{e3.5} has a solution $q$ with $(Q)$ for $F:=F[p]$ from Lemma {\rm\ref{l2.2}} and\\
$F_0(h):=F(h+E_0)$.
    \end{theorem}
    \textbf{Proof.}
Necessity: If $p$ is extendable, then there exists $q$ with $(Q)$ such that
    \begin{align*}
f(r,u)&=q\left(-U(r)-E_0+\frac{u^2}2\right), &(V)&\qquad \text{with}
    \\
U(r)&=-Lp(r),\quad E_0=Lp(R), &(P)& ,
    \\
p(r)&=\int_{\mathbb{R}^3} f(r,u)\,dv. &(D)&
    \end{align*}
Lemma \ref{l3.2} part (a) shows then that Eddington's equation \eqref{e3.5} has the
solution $q$ with $(Q)$.

Sufficiency: If Eddington's equation \eqref{e3.5} has a solution $q$ with $(Q)$ for
$F:=F[p]$, then $f:=f_q$ satisfies $(V)$ with $U(r):=-Lp(r)$ $(P)$ and $E_0=Lp(R)$.
Therefore, by virtue of $(V)$, \eqref{e3.1}, and \eqref{e3.3} we have
    \begin{multline*}
\int_{\mathbb{R}^3}\!\!f_q\big(r,|v|\big)\, dv=
\int_{\mathbb{R}^3}\!\!q\left(Lp(r)-E_0-\frac{v^2}2\right)dv
    \\
=4\pi\sqrt2\int_0^{Lp(r)-E_0}\!\!\!q(s) \sqrt{Lp(r)-E_0-s}\, ds
=F\circ Lp(r)=p(r),
    \end{multline*}
that is, $(D)$ is also valid, and $p$ is extendable.\hfill$\square$

In the next theorem, we investigate the solvability of Eddington's equation in the form
    \[
F_0(h)=4\pi\sqrt2\int_0^h q(s)\sqrt{h-s}\, ds
    \]
for given $F_0(h):=F(h+E_0)$, $F:=F[p]$, in more detail.

The following theorem gives different conditions of extendability for a function
$p\in\mathcal{D}_R^-(L)$ in explicit form. The spaces $L_{\rm
loc}^1[0,T]$ and $AC[0,T)$  are defined in the Appendix.
    \begin{theorem}\label{t4.2}
Let $p\in\mathcal{D}_R^-(L)${\rm,} \; $p\,\big|_{(0,R]}\in C^2(0,R]${\rm,} \;   $F:=F[p]${\rm,} \;
$E_0:=P(R)$ \; and \\ $F_0(\,\cdot\,)=F(\,\cdot\,+E_0)$. Then the following statements hold.
\\[1ex]
{\rm1)}~Eddington's equation
    \begin{equation}\label{e4.1}
F_0(h)=4\pi\sqrt2\int_0^h\!q(s)\sqrt{h-s}\,ds, \qquad 0\leq h<P(0)-E_0
    \end{equation}
has a unique real-valued solution $q\in L_{\rm loc}^1[0,P(0)-E_0)${\rm,} which is given
by
    \[
q(h):=\frac1{4\pi\sqrt2}\,\frac2\pi\,\frac d{dh}\, H_{F'_0}(h), \qquad 0\leq h<P(0)-E_0,
    \]
where
    \begin{equation}\label{e4.2}
H_{F_0'}(h):=\int_0^h\frac{F_0'(s)}{\sqrt{h-s}}\,ds \;\;\;\text{lies in}\;\; AC\big[0, P(0)-E_0\big), \quad
F_0\in C^2\big[0, P(0)-E_0\big).
    \end{equation}
{\rm 2)}~$p$ is extendable if and only if $q>0$ on $(0, P(0)-E_0)${\rm,} that is
    \begin{equation}\label{e4.3}
    \begin{gathered}
\frac d{dh}\,H_{F_0'}(h)=\frac1{\sqrt
h}\,F_0'(0)+\int_0^h\frac{F_0''(s)}{\sqrt{h-s}}\,ds>0 \quad \text{on} \quad \big(0,
P(0)-E_0\big)
    \\[0ex]
    \end{gathered}
    \end{equation}
$\left(F_0'(0)=\frac{p'(R)}{P'(R)}\ge0\right)$.
\\[2ex]
{\rm 3)}~Sufficient conditions for the extendability of $p$ are{\rm:}
    \begin{itemize}
\item[\rm (a)] $F_0''(s)>0$ on $\big(0, P(0)-E_0\big)${\rm,}
     \\
\item[\rm (b)] $X(r):=p'(r)\cdot P''(r)-p''(r)P'(r)>0$ on $(0,R)${\rm,}
    \\
\item[\rm (c)] $\dfrac2r\,p'(r)+p''(r)>0$ on $(0,R)$, 
    \end{itemize}
 where {\rm (}a{\rm)} and
{\rm (}b{\rm)} are equivalent and {\rm (}c{\rm)} implies {\rm (}a{\rm)} and {\rm(}b{\rm)}.
    %
    \end{theorem}
    \textbf{Proof.}
The assumptions on $p$ and Lemma \ref{l2.1} imply that
    \[
P\colon(0,R]\to[E_0,P(0))
    \]
is a strictly decreasing bijection in $C^2(0,R]$ with strictly decreasing inverse
    \[
P^{-1}\colon [E_0, P(0))\to(0,R]
    \]
in $C^2[E_0,P(0))$. The composition with $p\in C^2(0,R]$:
    \[
F:=p\circ P^{-1}\colon [E_0,P(0))\to [0,p(0))
    \]
is strictly increasing and $F\in C^2[E_0,P(0))$, $F_0\in C^2[0, P(0)-E_0)$,
$F_0'(\,\cdot\,)=F'(\,\cdot\,+E_0)$, and $F_0''(\,\cdot\,)=F''(\,\cdot\,+E_0)$.\\[-1ex]

1)~To show that \eqref{e4.1} has a unique real-valued solution $q\in L_{\rm loc}^1[0,
P(0)-E_0)$, we need to verify that for $g:=\dfrac{F_0}{4\pi\sqrt2}$ the assumptions (a), (i) and (ii) 
of Lemma \ref{lA.4} (in the Appendix) are satisfied. It is sufficient to do this
for $g:=F_0$.

Obviously, $F_0\in C^2[0, P(0)-E_0)\subset AC[0,P(0)-E_0)$ and
    \[
F_0(0)=F(E_0)=pP^{-1}(E_0)=p(R)=0.
    \]
For
    \[
H_{F_0'}(h):=\int_0^h\frac{F_0'(s)}{\sqrt{h-s}}\, ds
    \]
 (a) (i) means that we have to show  $H_{F_0'}\in AC[0, P(0)-E_0)$. We observe that we
have $F_0'\in C^1[0, P(0)-E_0)$ and that
    \begin{equation}\label{e4.4}
F_0'(0)=F'(E_0)=p'\big(P^{-1}(E_0)\big)\cdot (P^{-1})'(E_0)= \frac{p'(R)}{P'(R)}\ge0.
    \end{equation}
Integrating by parts, we get
    \begin{align*}
H_{F_0'}(h)&=-2\sqrt{h-s}\, F_0'(s)\big|_0^h+2\int_0^h\!\!F_0{''}(s)\sqrt{h-s}\, ds
    \\
&= 2\sqrt{h}\, F_0'(0)+2\int_0^h\!\!F_0''(s)\sqrt{h-s}\,ds,
    \end{align*}
and $H_{F_0'}\in AC[0,P(0)-E_0)$ follows, i.e. (i) is satisfied. Also we get $H_{F_0'}(0)=0$, which is (a) (ii).
\,It follows from Lemma~A.5 (a) that \eqref{e4.1} has a unique realvalued solution $q$
which is given by
    \begin{equation}\label{e4.5}
q(h):=\frac1{4\pi\sqrt2}\,\frac2\pi\,\frac{d}{dh}\, H_{F_0'}(h)= \frac1{4\pi\sqrt2}\,
\frac2\pi\left[\frac1{\sqrt{h}}\, F_0'(0)+\int_0^h\!\!\frac{F_0''(s)}{\sqrt{h-s}}\,
ds\right].
    \end{equation}
\\[-2ex]

2)~Since $H_{F_0'}\in AC\big[0, P(0)-E_0\big)$, we have  $q\in L_{\rm loc}^1[0, P(0)-E_0)$.
By\\ Theorem \ref{e4.1}, it satisfies $(Q)$ and $p$ is extendable if and only if $q>0$ on
$(0, P(0)-E_0)$.
\\[-1ex]

\begin{figure}[ht]\centering
   \caption{ Action of $\Phi$ and $\Psi$}\label{D43}
        \begin{minipage}{85mm}
    \centering\includegraphics[scale=0.68]{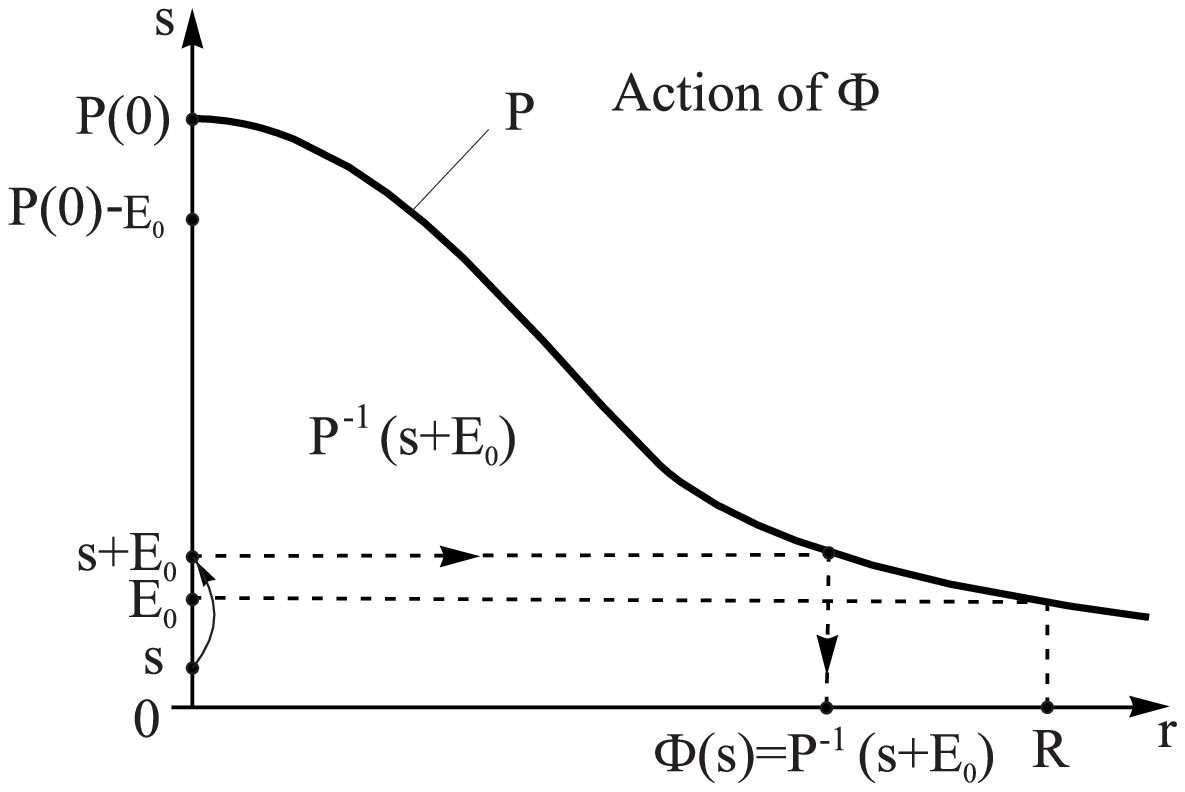}
    {
   \small For $0<s<P(0)-E_0$, we have  $E_0<s+E_0<P(0)$.
    Therefore $P^{-1}(s+E_0)=\Phi(s)<R$.}
    \vspace*{3mm}
    \end{minipage}
    \\[3mm]
    \begin{minipage}{80mm}
    \hspace*{-15mm}
    \centering\includegraphics[scale=0.64]{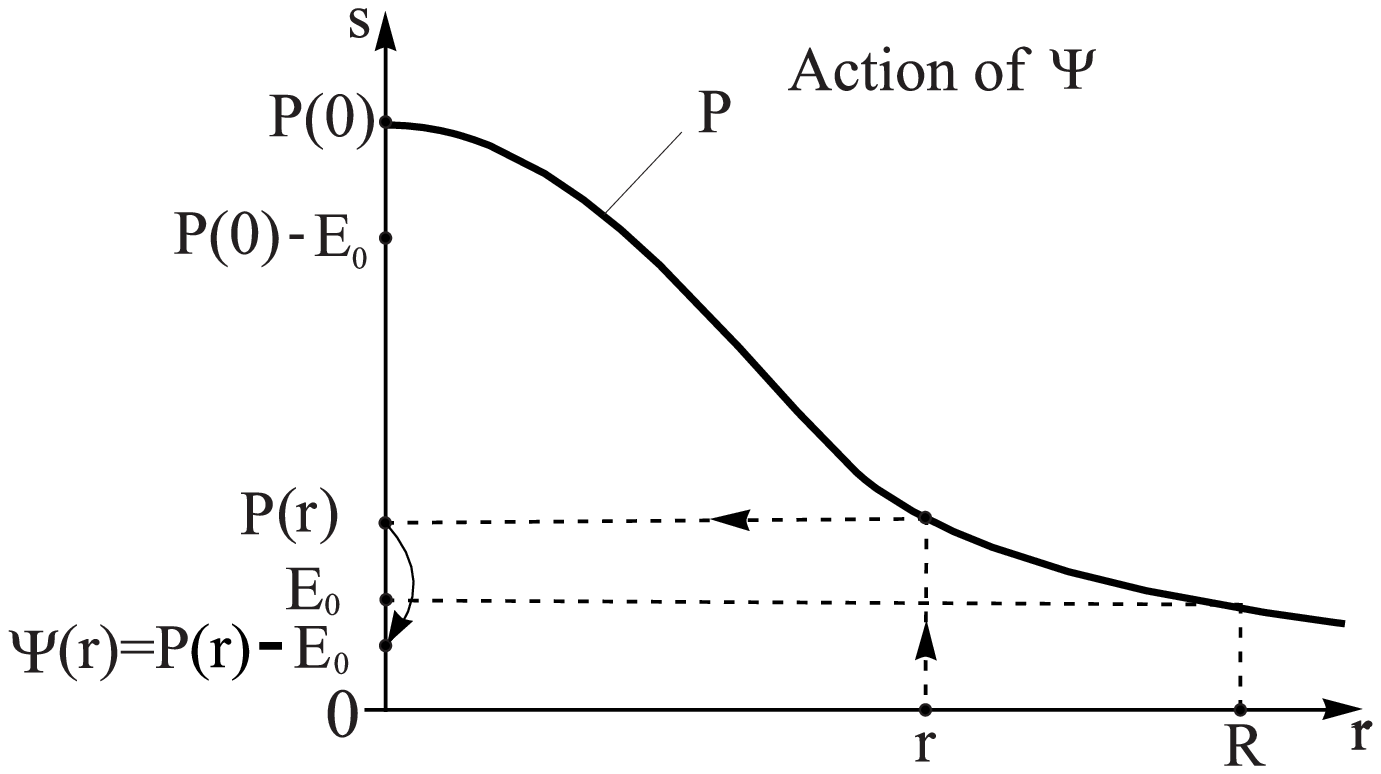}
    {\hspace*{0mm}
    \small \\For $0<r<R$, we have  $E_0<P(r)<P(0)$.
    Hence $0<P(r)-E_0=\Psi(r)<P(0)-E_0$.}
    \end{minipage}
        \end{figure}

3)~The proof of 3) is based on a change of variables in the integral
    \[
\int_0^h\!\!\frac{F_0''(s)}{\sqrt{h-s}}\, ds.
    \]
We define a $C^2$-diffeomorphism $\Phi$
    \[
\Phi\colon \big[0, P(0)-E_0\big)\to(0,R]
    \]
as the composition of the shift
    \[
T\colon\big[0, P(0)-E_0\big)\to\big[E_0, P(0)\big), \qquad s\mapsto s+E_0,
    \]
with
    \[
P^{-1}\colon \big[E_0, P(0)\big)\to(0,R], \qquad s\mapsto P^{-1}(s),
    \]
that is, $\Phi:=P^{-1}\circ T$, $s\mapsto r=P^{-1}(s+E_0)$ (see Diagram \ref{D43}).
%
The inverse of $\Phi$ is $\Psi:=T^{-1}\circ P$:
    \[
\Psi\colon(0,R]\to\big[0, P(0)-E_0\big), \qquad r\mapsto s=P(r)-E_0.
    \]
We now represent $F_0'(s)$, $F_0''(s)$ as functions of $p'(r)$, $p''(r)$, $P'(r)$,
$P''(r)$ as follows: for $s=\Psi(r)$, we have
    \begin{align*}
&F_0(s)=F_0\big(\Psi(r)\big)=F\big(P(r)-E_0+E_0\big)=F\big(P(r)\big)=pP^{-1}\big(P(r)\big)=p(r),
    \\
&\frac{d}{dr}\, F_0\big(\Psi(r)\big)=F'\big(P(r)\big)\cdot P'(r)=p'(r).
    \end{align*}
Since $P'(r)<0$, we obtain for $s=\Psi(r)$
    \begin{align*}
F_0'(s)&=F_0'\big(\Psi(r)\big)=F'\big(P(r)\big)=\frac{p\,'(r)}{P'(r)},
    \\
\frac{d}{dr}\,F_0'\big(\Psi(r)\big)&=\frac{d}{dr}\,F'\big(P(r)-E_0+E_0\big)=\frac{d}{dr}
F'\big(P(r)\big)
    \\
&= F''\big(P(r)\big)\cdot P'(r)=\frac{d}{dr}\,\frac{p\,'(r)}{P'(r)}
    \\
&=\frac{P'(r)\cdotp\,p\,''(r)-p\,'(r)\cdot P''(r)}{P'(r)^2}.
    \end{align*}
Hence, for $s=\Psi(r)$, we have
    \[
F_0''(s)=F_0''\big(\Psi(r)\big)=F''\big(P(r)\big)=\frac{p\,'(r)
P''(r)-p\,''(r)P'(r)}{\big|P'(r)\big|^3}\,.
    \]
Changing coordinates $s\to r$ in the integral of \eqref{e4.3} by
    $\Phi(s):=P^{-1}\circ T(s)=P^{-1}(s+E_0)=r$,
    $\Phi(0)=R$,
    $\Phi(P(0)-E_0)=0$,
    $\Phi(h)=P^{-1}\circ T(h)=P^{-1}(h+E_0)$,
    $ds=P'(r)\,dr$, and observing the negative sign of $P'(r)$, we get
    \begin{equation}\label{e4.6}
\int_0^h\!\!\frac{F_0''(s)}{\sqrt{h-s}}\, ds=\int_{P^{-1}(h+E_0)}^R\!\frac
{p\,'(r)P''(r)-p\,''(r)P'(r)}{\big|P'(r)\big|^3\sqrt{h-\big(P(r)-E_0\big)}}\,
\big|P'(r)\big|\,dr,
    \end{equation}
and 3) (a) or 3) (b) imply $q>0$ in view of \eqref{e4.3}, \eqref{e4.4}.

Let 3) (c) hold. Then, by virtue of Lemma \ref{l2.2}, we have
    \begin{align*}
p\,'P''-p\,''P'&=p\,'\left(-\frac2r\,P'-4\pi p\right)-p\,''P'
    \\
&=4\pi p(-p\,')+(-P') \left(\frac 2r \,p\,'+p\,''\right)>0,
    \end{align*}
and 3) (b) is fulfilled.\hfill$\square$
    %
    %

    \begin{lremark}\label{r4.1}
For later numerical calculations it is useful to write \eqref{e4.3} in another form.
Because $\Phi(h)=P^{-1}(h+E_0)$, we have $P\big(\Phi(h)\big)=h+E_0$ and therefore
    \[
h=P\big(\Phi(h)\big)-P(R),
    \]
and with \eqref{e4.4} and \eqref{e4.6} we get for \eqref{e4.3}
    \begin{multline}\label{e4.7}
\frac{d}{dh}\,H_{F_0'}(h) =\frac{1}{\sqrt{P\big(\Phi(h)\big)-P(R)}}
    \\
\times\left(\frac{p\,'(R)}{P'(R)}+\int_{\Phi(h)}^R\!\!\frac{p\,'(r)P''(r)-p\,''(r)P'(r)}
{\big|P'(r)\big|^2}\sqrt{\frac{P\big(\Phi(h)\big)\!-\!
P(R)}{P\big(\Phi(h)\big)\!-\!P(r)}}\, dr\right)
    \end{multline}
The integrand has a singularity at $r=\Phi(h)$.
    \end{lremark}
    \begin{lremark}\label{r4.2}
Examples in the following section will show that the conditions 3) (a), (b), and (c) are
not necessary for the extendability.
    \end{lremark}
$~$\\[-9ex]

\section{ Examples}\label{s5}
%
%
    \begin{lexample}\label{ex5.1}
    \[
p(r):=\begin{cases}
    1-\left(\dfrac{r}{R}\right)^2\!,  &\quad 0\le r\le R,
        \\
    0, &\quad R< r, \;   R>0.
        \end{cases}
    \]
This example allows to compute explicitly the other functions involved in the theory:
$P$, $P^{-1}$, $F$, $q$, $G_0=:F_0^{-1}$ (which is the right hand side of the
approximating nonlinear system \eqref{eANS} occurring in Section \ref{s7}).

For $0\le r\leq R$, we have
    \[
p\,'(r)=-\frac2{R^2}\,r, \qquad p\,''(r)=-\frac2{R^2},
    \]
hence $\dfrac2r\, p\,'(r)+p\,''=-\dfrac6{R^2}<0$. Therefore the sufficient condition 3) (c)
in Theorem \ref{t4.2} does not hold.

Substituting $p$ into \eqref{e2.1}, we obtain
    \begin{allowdisplaybreaks}
    \begin{align*}
P(r)&=4\pi\left[\frac1r\int_0^r\left(1-\frac{s^2}{R^2}\right)s^2\,ds+\int_r^R\!
\left(1-\frac{s^2}{R^2}\right)s\,ds\right]
    \\
&=4\pi\left[\frac{r^2}3-\frac15\,\frac{r^4}{R^2}+\frac12\big(R^2-r^2\big) -\frac14\,
\frac1{R^2}\big(R^4-r^4\big)\right]
    \\
&=4\pi R^2\left[\frac1{20}\left(\frac{r}{R}\right)^4\!-\!\frac16\left(\frac{r}R\right)^2
\!+\!\frac14\right] =\pi R^2\left[\frac15\left(\frac{r}R\right)^4-
\frac23\left(\frac{r}R\right)^2\!+\!1\right]\!:=\! h,
    \\
P'(r)&=4\pi R^2\left[\frac15\,\frac{r^3}{R^4}-\frac13\,\frac{r}{R^2}\right],
    \\
P''(r)&=4\pi R^2\left[\frac35\,\frac{r^2}{R^4}-\frac13\,\frac1{R^2}\right],
    \\
P(R)&=E_0=\frac8{15}\,\pi R^2
    \end{align*}
    \end{allowdisplaybreaks}
and it follows
    \begin{align*}
X(r)&=p\,'P''-p\,''P'=-\frac2{R^2}\,r\cdot4\pi R^2\left[\frac35\, \frac{r^2}{R^4}-
\frac13\, \frac1{R^2}\right]
    \\
&\qquad\qquad\qquad\quad\;\;+ \frac2{R^2}\cdot 4\pi R^2\left[\frac15\, \frac{r^3}{R^4}-\frac13\,
\frac{r}{R^2}\right]
=-\frac{16\pi}5\,\frac{r^3}{R^4}<0,
    \end{align*}
and sufficient condition~3) (b) in Theorem \ref{t4.2} is not fulfilled.

Since $P(r)=h$ and $P\colon[0,R]\to\big[E_0, P(0)\big]=\Big[\dfrac8{15}\,\pi R^2, \pi
R^2\Big]$ is a biquadratic form in $\dfrac{r}{R}$, we can calculate $P^{-1}$:
    \begin{align*}
\left(\frac{r}{R}\right)^2&=\frac53-\sqrt{\frac{25}9-5+\frac5{\pi R^2}\,h}\,,
    \\
r&=P^{-1}(h)=R\sqrt{\frac53-\sqrt{\frac5{\pi R^2}\,h-\frac{20}9}}\,.
    \end{align*}
Hence we obtain
    \begin{align*}
F(h)&=p( P^{-1}(h))=1-\left(\frac{P^{-1}(h)}R\right)^2=1-\left(\frac53-
\sqrt{\frac5{\pi R^2}\,h-\frac{20}9}\,\right)
    \\
&=\sqrt{\frac5{\pi R^2}\,h-\frac{20}3}-\frac23\,, \qquad h\in\big[E_0, P(0)\big],
    \\
F_0(h)&=F(h+E_0)=\sqrt{\frac5{\pi R^2}\left(h+\frac8{15}\,\pi R^2\right)-\frac{20}9}
-\frac23
    \\
&=\sqrt{ah+\frac49}-\frac23\,, \quad \text{with} \quad a:=\frac5{\pi R^2}, \
h\in\big[0,P(0)-E_0\big].
    \end{align*}
We have
    \begin{align*}
    &F_0(0)=0,\qquad
    F_0(P(0)-E_0)=F_0\Big(\dfrac7{15}\,\pi R^2\Big)= \sqrt{\dfrac73+
\dfrac49}-\dfrac23\,=1,
    \\
    &F_0'(h)=\dfrac{a}2\dfrac1{\sqrt{ah+\dfrac49}},
        \\
    &F_0'(0)=\dfrac34\,a,\qquad
    F_0'(P(0)-E_0)=\dfrac{3a}{10},
        \\
    &F_0''(h)=-\dfrac{a^2}4\dfrac1{\Big(\sqrt{ah+\dfrac49}\Big)^3}.
    \end{align*}

From \cite[p. 306]{4} it follows that
    \begin{align*}
H_{F_0'}(h)&=\int_0^h\!\!\frac{F_0'(s)}{\sqrt{h-s}}\,ds=
\frac{a}2\int_0^h\!\!\frac{ds}{\sqrt{as+\dfrac49}\,\sqrt{h-s}}
    \\
&= \frac{a}2\,\frac2{\sqrt{a}}\,\mathop{\rm arctg}\sqrt{
\dfrac{(as+4/9)}{a(h-s)}}\,\Biggl|_{s=0}^{s=h}
=\sqrt{a}\left(\frac{\pi}2-\mathop{\rm arctg}\left(\frac23\,
\frac1{\sqrt{a}}\dfrac1{\sqrt{h}}\right)\right).
    \end{align*}

Differentiating the last expression, we obtain
    \begin{align*}
\frac{d}{dh}\,H_{F_0'}(h)&=-\frac{\sqrt{a}}{1+\dfrac49\, \dfrac1{a}\,
\dfrac1h}\,\frac23\cdot\frac1{\sqrt{a}}\,\Big(-\frac12\Big)\,\frac1{h^{3/2}}
   \\
&=\frac13\,\dfrac1{h+\dfrac49\,\dfrac1a}\cdot \frac1{\sqrt{h}}>0, \qquad    h\in(0,P(0)-E_0).
    \end{align*}
We conclude that $H_{F_0'}$ is in $AC\big[0,P(0)-E_0\big)$, $H_{F_0'}(0)=0$ and
$\dfrac{d}{dh} H_{F_0'}(h)>0$ on $(0, P(0)-E_0)$. Hence, by virtue of Theorem \ref{t4.2}\; 1) and
\ref{t4.2}\; 2), $p$ is extendable and
    \[
q(h):=\frac1{4\pi\sqrt{2}}\,\frac2\pi\, \frac{d}{dh}\, H_{F_0'}(h)=
\frac{\sqrt{2}}{4\pi^2}\, \frac13\,\frac1{h+\dfrac49\,\dfrac1a}\cdot\frac1{\sqrt{h}}>0
    \]
is the unique solution of Eddington's equation \eqref{e4.1}. 

This example illustrates as well the inverse problem (with known $p$) as the direct
problem (with known $q$ --- in Section \ref{s6}). It will be expanded further in Section
\ref{s8}. 
\\[-2ex]
    \end{lexample}
    %
%
    \begin{lexample}\label{ex5.2}
    \[
p(r):=\begin{cases}
        \Big(1-\dfrac{r}{R}\Big)^2, \ &0\le r\le R,
            \\
        0, \ & R< r, \;  R>0.
        \end{cases}
    \]
This is an example, which allows to decide about its extendability easily by means of
Theorem \ref{t4.2}\; 3) (b). For $0\le r\le R$, we have
    \[
p\,'(r)=-\frac2R+2\,\frac{r}{R^2}, \qquad p\,''(r)=\frac2{R^2}\,.
    \]
The function $\dfrac2r\,p\,'(r)+p\,''(r)=\dfrac6{R^2}-\dfrac4{rR}$ changes its sign at
the point $\dfrac{2R}3\in(0,R)$. Thus Theorem \ref{t4.2}\; 3) (c) is not applicable. From
\eqref{e2.1} we obtain
    \begin{allowdisplaybreaks}
    \begin{align*}
P(r)&=4\pi\left[\frac1r\int_0^r\!\!\left(s^2-\frac{2s^3}R+\frac{s^4}{R^2}\right) ds
+\int_r^R\!\!\left(s-\frac{2s^2}R+\frac{s^3}{R^2}\right) ds\right]
    \\
&=4\pi\left[\frac13\,r^2\!-\!\frac12\,\frac{r^3}R\!+\!\frac15\,\frac{r^4}{R^2}\!+\!\frac12
\big(R^2\!-\!r^2\big)\!-\!\frac23\,\frac1R\big(R^3\!-\!r^3\big)+\frac14\,\frac1{R^2}
\big(R^4\!-\!r^4\big)\right]
    \\
&=4\pi\left[-\frac16\,r^2+\frac16\,\frac{r^3}R-\frac1{20}\,\frac{r^4}{R^2}+
\frac{3}{4}\,R^2\right],
    \\[0.3ex]
P'(r)&=4\pi\left[-\frac13\,r+\frac12\,\frac{r^2}R-\frac15\,\frac{r^3}{R^2}\right],
    \\[0.3ex]
P''(r)&=4\pi\left[-\frac13+\frac{r}R-\frac35\,\frac{r^2}{R^2}\right],
    \\[2ex]
X(r)&=p\,'(r)P''(r)-p\,''(r)P'(r)
    \\
&=4\pi\left[\left(-\frac2R\!+\!2\frac{r}{R^2}\right)
\left(-\frac13\!+\!\frac{r}R-\frac35\,\frac{r^2}{R^2}\right)\!-\!\frac2{R^2}\left(-\frac13\,
r+\frac12\,\frac{r^2}R-\frac15\,\frac{r^3}{R^2}\right)\right]
    \\
&=\frac{4\pi}R\left(\frac23-2\alpha+\frac{11}5\,\alpha^2-\frac45\,\alpha^3\right)=:
\frac{4\pi}Rf(\alpha) \ \text{with} \ \alpha:=\frac{r}R\leq1.
    \end{align*}
    \end{allowdisplaybreaks}
On $[0,1]$ we have $f'(\alpha)=-2+\dfrac{22}5\,\alpha-\dfrac{12}5\,\alpha^2$. It is easy
to see that $f$ is decreasing on $\Big[0,\dfrac56\Big]$, increasing on
$\Big[\dfrac56,1\Big]$, and $f(5/6)=\dfrac7{108}$. Hence $f(\alpha)>0$ on $[0,1]$ and
$X(r)>0$ on $(0,R)$. Hence, by virtue of Theorem \ref{t4.2}\; 3) (b) $p$ is extendable.

In this case it is not possible to calculate $P^{-1}$ explicitly (as in Example
\ref{ex5.1}), because $P(r)$ is a monotone non special polynomial of degree 4.
    \end{lexample}.
$~$\\[-5ex]  
    %
    \begin{lexample}\label{ex5.3}
    \[
p(r):=\begin{cases}
        e^{-r}-e^{-R}, \ &0\leq r\leq R
            \\
        0, \ & R< r, \;  R>0.
        \end{cases}
    \]

\noindent We have for $0<r\le R$
    \begin{align*}
p\,'(r)&{\phantom{:}}=-e^{-r}, \qquad p\,''(r)=e^{-r},
    \\
P(r)& :=4\pi\left[\frac1r\int_0^r\!\!\big(e^{-s}-e^{-R}\big)s^2\,ds+
\int_r^R\!\!\big(e^{-s}-e^{-R}\big)s\,ds \right]
    \\
&{\phantom{:}}=4\pi\left[\frac1r\left(e^{-s}\big(-s^2-2s-2\big)-e^{-R}\frac{s^3}3\right)\bigg|_0^r
+\left(e^{-s}(-s-1)-e^{-R}\frac{s^2}2\right)\bigg|_r^R\,\right]
    \\
&{\phantom{:}}=4\pi\left[-e^{-r}\left(r+2+\frac2r\right)+\frac2r-\frac13\,
r^2e^{-R}-e^{-R}\left(R+1+\frac{R^2}2\right) \right.\\
&\left.\;\;\;\;\;\;\;\;\;\;\;\;\;\; +e^{-r}(r+1)+e^{-R}\,\frac{r^2}2\right]
    \\
&{\phantom{:}}=4\pi\left[-e^{-r}\left(1+\frac2r\right)+\frac2r-e^{-R}
\left(1+R+\frac{R^2}2-\frac{r^2}6\right)\right],
    \\
P'(r)&{\phantom{:}}=4\pi\left[e^{-r}\left(1+\frac2r+\frac2{r^2}\right)
-\frac2{r^2}+\frac13\,e^{-R}r\right],
    \\
P''(r)&{\phantom{:}}=4\pi\left[-e^{-r}\left(1+\frac2r+\frac4{r^2}
+\frac4{r^3}\right)+\frac4{r^3}+\frac13\,e^{-R}\right].
    \end{align*}
We get on the triangle $\{ (r,R); \; 0 < R,\; r \in (0,R]\}$  \\[-2ex]
    \begin{align*}
X(r,R)&=p\,'(r)P''(r)-p\,''(r)P'(r)=-e^{-r}\Big(P'(r)+P''(r)\big)
    \\
&=-4\pi e^{-r}\left[e^{-r}\left(1+\frac2r+\frac2{r^2}\right)-\frac2{r^2}
+\frac13\,e^{-R}r-e^{-r}\left(1+\frac2r+\frac4{r^2}+\frac4{r^3}\right)\right. \\
& \left. \;\;\;\;\;\;\;\;\;\;\;\;\;\;\;\;\;\;\;\;\;\;+\frac4{r^3}+\frac13\,e^{-R}\right]
    \\
&=4\pi e^{-r}\left[ \Big\{ e^{-r}\left(\frac2{r^2}+\frac4{r^3}\right)+\frac2{r^2} \Big\} 
-\Big\{ \frac4{r^3}+\frac{e^{-R}}3\,(r+1) \Big\}\right]
    \\
&=\;\;\;4\pi\,\frac{e^{-r}}{r^3}\left[ \{e^{-r}(2r+4)+2r\} - \{4+\frac13\,e^{-R}(r^4+r^3)\} \right].
   \end{align*} 
  \\[-3ex]

Because \\[-1ex]
  \begin{equation}\label{e5.1}
\frac{\partial X}{\partial R} (r,R)  =\frac{4\pi}{3}\; \frac{r+1}{e^{r+R}} > 0  \;\;\;\; \text{for} \;\; 0<R, \;\;  r\in (0,R]   
  \end{equation}
\\[0ex]  
X is  strictly increasing in direction of growing values of  $R > 0$ and each $r$ fixed in $\big[0 ,  R\big]$.
 \\[3ex]
It seems that $X, P, P', P''$ have singularities at $r=0$. But they can be repaired, as shown for $X$ in (\ref{e5.2}) and (\ref{e5.3}).
 For $ P, P', P''$
it can be done in the same way.  
The last formula of $X(r,R)$ is not useful to calculate values of  $X(r,R)$ with $r$ near zero, because X is the difference of
two large, positive terms, who have nearly equal values. Therefore we make 
the following transformation: \\[1ex]
If $\;\;\;\tilde{X}(r,R):= e^{2r} r^3 (4\pi)^{-1}\cdot X \left(r,R\right),\;\;$ then we get 
  \begin{align*}
\tilde{X} \left(r,R\right)  & =\left[ (2r+4) -\frac{e^{r-R}}{3} r^3 (1+r) + (2r-4) e^r \right] 
\\
& =\left[ -\frac{e^{r-R}}{3} r^3 (1+r) + ( 2r+4) + (2r-4) \left(1+r+ \frac{r^2}{2!}+ \frac{r^3}{3!}
+ \frac{r^4}{4!}+ \sum_{k=5}^{\infty} \frac{r^k}{k!} \right)\right]
\\  
&=\left[ -\frac{e^{r-R}}{3} r^3 (1+r)+ (2r+4)+( 2r+2r^2+ \frac{2r^3}{2!}+ \frac{2r^4}{3!}+ \frac{2r^5}{4!} ) \right. \\
&\left.\;\;\;\;\;\;\; + (-4-4r- \frac{4r^2}{2!}- \frac{4r^3}{3!}- \frac{4r^4}{4!}) + (2r-4) \sum_{k=5}^{\infty} \frac{r^k}{k!}  \right]
\\  
&=\left[ -\frac{e^{r-R}}{3} r^3 (1+r) +r^3 ( 1-\frac{4}{6}) + r^4 ( \frac{2}{6}- \frac{1}{6})
+ r^5\frac{2}{4!}+ (2r-4) \sum_{k=5}^{\infty} \frac{r^k}{k!} \right] 
\\ 
& = r^3 \left[ -\frac{e^{r-R}}{3} (1+r) + \frac{1}{3} + \frac{r}{6}+ \frac{r^2}{12}
+ (2r-4) r^2  \sum_{k=5}^{\infty} \frac{r^{k-5}}{k!} \right].  
 \end{align*}\\[-1ex]
Dividing $\tilde{X}$ by \;$e^{2r} r^3 (4\pi)^{-1}$\; we have \\[-1ex]
  \begin{equation}\label{e5.2}
X(r,R)=\frac{4\pi}{e^{2r}}  \left[ -\frac{e^{r-R}}{3} (1+r) + \frac{1}{3} + \frac{r}{6}+ \frac{r^2}{12}
+ (2r-4) r^2  \sum_{k=5}^{\infty} \frac{r^{k-5}}{k!} \right].
  \end{equation}
It follows 
  \begin{equation}\label{e5.3}
 \lim\limits_{r\to0} X(r,R)=
 4\pi \left[  -\frac{1}{3 e^R} + \frac{1}{3} \right] \in \left(0,\frac{4\pi}{3}\right] \;\;\; \text{for} \; R>0.   
  \end{equation}
 \\[-2ex]

First we calculate $\; X(r,r)$
\\ [-3ex]     
   \begin{align} \label{e5.4}
 X(r,r)&=\frac{4\pi}{e^{2r}}  \left[ -\frac{e^{r-r}}{3} (1+r) + \frac{1}{3} + \frac{r}{6}+ \frac{r^2}{12}
 + (2r-4) r^2  \sum_{k=5}^{\infty} \frac{r^{k-5}}{k!} \right]\nonumber
\\ 
&=\frac{4\pi}{e^{2r}} r  \left[   \frac{r}{12} - \frac{1}{6} + (2r-4)  \sum_{k=5}^{\infty} \frac{r^{k-4}}{k!} \right] \nonumber
\\
&  \begin{cases}
=0 & \text{\quad for }r=2, \\
>0 & \text{\quad for }2<r,\\ 
<0 & \text{\quad for }r<2. 
\end{cases} 
 \end{align}\\[-1ex]

A possible application of  Theorem \ref{t4.2} 3) (b) requires the determination of the sign of $X(r,R)$ for $0<r<R$.
We distiguish between two cases
\[
(1)\; R\ge 2 \text{\quad and \quad} (2)\; R < 2
\]

In case (1) and for $0<r<2$ we know from  (\ref{e5.1}) and Chart (\ref{tab5.7}) (Diagram \ref{fig5.8}),  that
\[ X(r,R)>X(r,2) > 0,\]
for $2 \le r < R$ we have with (\ref{e5.1}) and  (\ref{e5.4}) that  
\[ X(r,R) > X(r,r) \ge 0. \]
It follows that for \;case (1) $R\ge 2$\; we have $X(r,R)>0$ \; for\; $0<r<R$ \; and p is extendable by  Theorem \ref{t4.2} 3) (b).\\[-1.5ex] 

 In the case (2)\; with $0<  R < 2$, we know from (\ref{e5.4}) that\;  $X(r,r) < 0$, on the other hand\; $X(r,2)>0$ for $r \in [0,2)$.
By (\ref{e5.1})  there exist a unique zero of $X(r,R)$ between $R$ and $2$ for each r fixed in $(0,2)$. The zeros ly on
a monotone curve produced by $X(r,R)=0$ in the neigborhood above the diagonal (see Diagram \ref{fig5.8}) :
\[
R(r)=-ln \left( 6 \frac{( r+2) e^{- r} + r- 2}{(r+ 1) r^3} \right) \;\; \text{for}\;\; 0<r\le R<2. 
\]
Therefore the  extendability of $p$ can be only decided with  Theorem \ref{t4.2} \;2) or 
 \eqref{e4.7} of Remark \ref{r4.2}.
 The integral is not calculable explicitly and the integrand is singular at the left end of
the integration interval. But the integral can be approximated by a Newton-Cotes formula 
avoiding border points (for instance the Midpoint Rule). 
The error occurring by using approximation formulas has to be
carefully estimated against the preceding term $p\,'(R)/P'(r)$ to claim the extendability
of $p$ by Theorem \ref{t4.2} \;2).\\
 Such a case will be treated in Example \ref{ex5.9} and we leave the details here to the reader.
     \renewcommand{\arraystretch}{1.5} 
\vspace{-6mm}
  \begin{table}[h!]
     \caption{~ $X(r,2))$ \; for\; $0\le r\le 2$} \label{tab5.7}
     \begin{center}
     \begin{tabular}{|c||c|c|c|c|c|c|c|c|c|c|}
     \hline
$r$ &$ 0 $&$10^{-5}$ &$0.25$&$0.5$ &$0.75$ &$ 1$ &$1.25$ &$1.5$&$1.75$ &$2$
   \\
    \hline     
$ X(r,2) $ &$3.6220$&$3.6218$&$2.33$ &$1.47$ &$0.910$&$0.541$ &$0.302$&$0.151$&$0.056$&$0$
    \\    
  \hline
     \end{tabular}
     \end{center}
     \end{table}   
$~$\\[-10ex]
   \begin{figure}[h!]\centering
 \begin{minipage}{100mm}\caption{ Ranges of positive and negative values of $X(r,R)$ }
    \label{fig5.8}
    \includegraphics[scale=0.21]{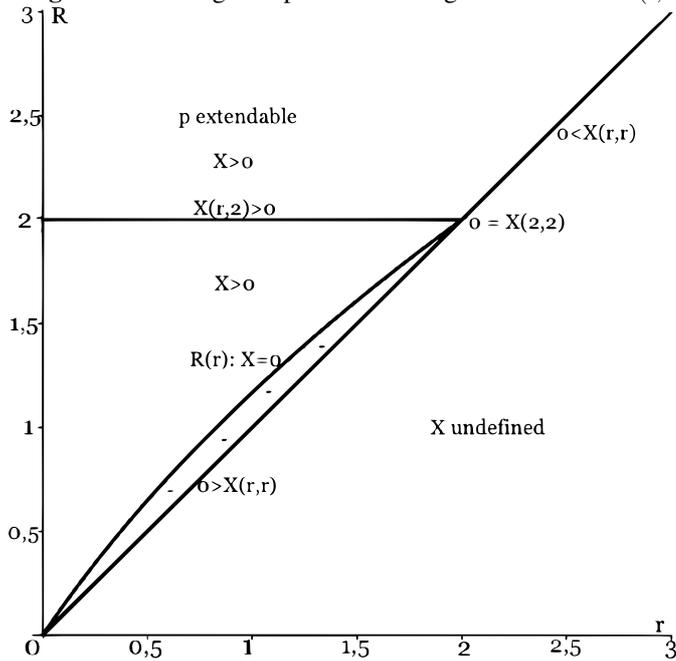}
    \end{minipage}
     \end{figure}
$~$\\[0ex]
\end{lexample}
%
%
     \begin{lexample}\label{ex5.4}
\[
p(r):=\begin{cases}
       r^{-b}-R^{-b}, \ &0< r\leq R
            \\
        0, \ & R < r, \;  R>0, \ 0<b<3.
        \end{cases}
    \]
First, we note that, if $b<3$, then $p(r)\in\mathcal{D}_R^-(L)$. \\[1ex]

For $0<r\le R$,we have
    \[
p\,'(r)=-br^{-b-1}, \qquad p\,''(r)=b(b+1)r^{-b-2}.
    \]
For $1<b<3$ we get
    \[
\frac2r\,p\,'(r)+p\,''(r)=-2br^{-b-2}+b(b+1)r^{-b-2}=b(b-1)r^{-b-2}>0,
    \]
and it follows from Theorem \ref{t4.2}\, 3) (c) that $p$ is extendable.
\\[2ex]
For $0<b<3$, $b\ne2$ and $0<r<R$, we have
    \begin{align*}
P(r)&=4\pi\left[\frac1r\int_0^r\!\!\big(s^{-b+2}-R^{-b}s^2\big)\,ds+ \int_r^R\!\!
\big(s^{-b+1}-R^{-b}s\big)\,ds\right]
    \\
&=4\pi\left[\frac1{-b+3}\,r^{-b+2}-\frac13\,R^{-b}r^2+\left(\frac1{-b+2}\,
s^{-b+2}-R^{-b}\frac{s^2}2\right)\bigg|_r^R\right]
    \\
&=4\pi\left[\frac{r^{-b+2}}{-b+3}-R^{-b}\frac{r^2}3+\frac{R^{-b+2}}{-b+2}
-\frac{r^{-b+2}}{-b+2}-\frac{R^{-b+2}}2+\frac{R^{-b}r^2}2\right]
    \\
&=4\pi\left[\frac{-r^{-b+2}}{(-b+3)(-b+2)}+\frac16\,R^{-b}r^2+ \frac{b}{2(-b+2)}\,
R^{-b+2}\right].
    \end{align*}
For $0<r<R$, we have
    \begin{align*}
\frac{X(r)}{4\pi}&=(-b)r^{-b\!-\!1}\left(\frac{b-1}{-b+3}\,r^{-b}\!+\!\frac13\,R^{-b}\right)\!-\!
b(b\!+\!1)r^{-b-2}\left(\frac{-r^{-b+1}}{-b+3}\!+\!\frac13\,R^{-b}r\right)
    \\
&=\frac{2b}{-b+3}\,r^{-2b-1}-\frac13(b^2+2b)R^{-b}r^{-b-1}
    \\
&=\frac{R^{-b}r^{-b-1}}{-b+3}
\left(2b\left(\frac{R}r\right)^b-\frac{b^2}3(3-b)-\frac{2b}3\,(3-b)\right)
    \\
&=\frac{R^{-b}r^{-b-1}}{-b+3}\cdot
b\cdot\left(2\left(\bigg(\frac{R}r\bigg)^b-1\right)+\frac13(b^2-b)\right).
    \end{align*}
For $1\le b<3, b\ne 2$, we get $X(r)>0$ and we conclude from Theorem \ref{t4.2}\, 3) (b) that $p$ is
extendable.\\
 For $0<b<1$, $X$ has a zero at $r_0$ on $(0,R)$, which is a solution of the
equation
    \[
\left(\frac{R}r\right)^b-1-\frac16\,(b-b^2)=0.
    \]
It follows
    \[
\frac{r_0}R=\left(\frac6{6+(b-b^2)}\right)^{\tfrac1b}.
    \]
We have $\dfrac{r_0}R=0,9216$ \ for $b=\dfrac12$, $\dfrac{r_0}R=1$ for $b=1$, and
$\lim\limits_{b\to 0}\dfrac{r_0}R\,(b)=e^{-1/6}\thickapprox 0,8465$.\\[1ex]
The existence of a zero of $X$ requires the application of  Theorem \ref{t4.2}\, 2). We omit the details.
\\[1ex]
     \end{lexample}

%
 %
     \begin{lexample}\label{ex5.8} $~$ \\[0ex]

This example is intended to illustrate the direct problem with given $q$ and $p$
calculated approximately in Section \ref{s8}.

Let $q(s):=c\sqrt{s}$, where $c>0$. Then, changing the variable $s=h\sin^2 t$, we have
\[
F_0(h):=4\pi\sqrt{2}\cdot c\!\int_0^h\!\!\!\sqrt{s}\, \sqrt{h-s}\,ds= \frac{\pi^2c}{\sqrt{2}}\, h^2.
\]
see \cite[p. 307]{4}.
     \end{lexample}
 $~$\\[-2ex]
     
    %
%
     \begin{lexample}\label{ex5.9} $~$ An unextendable $p$. \\[0ex]

The aim of this example is to show numerically that not all $p$, which satisfy the assumptions of Theorem \ref{t4.2}, 
are automatically extendable. That is, we construct a $p$ that will differ in important details from the extendable examples 
so far given: whereas all our examples $p$ where either convex or concave on $[0,R]$, this example will be concave in
in the subinterval $[0,w]$ and convex in  $[w,R]$ for some $0<w<R$ ($p''(w)=0$). Inequality (\ref{e4.3}) and formula (\ref{e4.7}) 
will play an important role.\\[-2ex] 

 We shall construct a function
\begin{equation*}
p(r)=\left\{\begin{array}{lc}
a_0+a_1r+a_2r^2+a_3r^3+a_4r^4, & r\in[0,R],\\
0, & r>R,
\end{array}\right.
\end{equation*}
such that $p\in{\cal D}_R^-(L)$ and $\dfrac{d}{dh}H_{F'_0}(h)<0$ for some $h\in(0,P(0)-E_0)$. Then, by virtue of
Theorem \ref{t4.2}\;2), $p$ will be not extendable.\\[-2ex]

The condition  $p\in{\cal D}_R^-(L)$ requires
\begin{align}
& p(0)>0,\label{e5.5}\\
& p(R)=0,\label{e5.6}\\
\hspace{-28mm} \text{We add the properties} \nonumber \\
& p'(0)=0,\label{e5.7}\\
& p'(R)=0,\label{e5.8}\\
& p''(w)=0\quad\text{for some}\, w\in(0,R).\label{e5.9}
\end{align}
\\[-2ex]
 The equalities \eqref{e5.8} and \eqref{e4.4} imply that $F'_0(0)=0$  
and this simplifies \eqref{e4.7}.
\\[1ex]
A moments' reflexion shows then that 
(5.5), (5.6), (5.7), (5.8), (5.9) hold if and only if \quad $a_0>0,  \;\; a_1=0$ \quad  and\\[-4ex] 
\begin{align}
&a_2R^2+a_3R^3+a_4R^4=-a_0,\label{e5.10}\\
&2a_2R+3a_3R^2+4a_4R^3=0,\label{e5.11}\\
&2a_2+6a_3w+12a_4w^2=0.\label{e5.12}
\end{align}$~$\\[-6ex]

Further we assume that 
\begin{equation}\label{e5.14}
R=2,\qquad w=\dfrac{13}{10},\qquad a_0=2.
\end{equation}\\[-2ex]
This choise of\; $R, w, a_0$\; cannot be made arbitrarily. Tests of different values of R show that for\; $w\le R/4$\; or\; $3R/4\le w$
\; $p$ will never be strictly monotone decreasing whatever $a_0>0$ may be. For\; $R/4 < w \le R/2$ \; 
 p is strictly  decreasing, but X is nowhere zero in $(0,R)$. 
For \;$R/2 < w < 3R/4$\;  $p$ is strictly decreasing and X has a zero in $(0,R)$.\\[-1ex]

The determinant of linear algebraic system \eqref{e5.10}--\eqref{e5.12} relatively to $a_2, a_3$, and $a_4$ is not zero. 
Therefore there exists exactly one solution of system \eqref{e5.10}--\eqref{e5.12}
$$
a_2=-\dfrac{39}{146},\qquad a_3=-\dfrac{107}{146},\qquad a_4=\dfrac{45}{146}.
$$

We now have
\begin{align}
& p(r)=2+\dfrac1{146}\left(-39r^2-107r^3+45r^4\right),\label{e5.15}\\
& p'(r)=\dfrac1{146}\left(-78r-321r^2+180r^3\right),\label{e5.16}\\
& p''(r)=\dfrac1{146}\left(-78-642r+540r^2\right),\label{e5.17}
\end{align}
(see Diagram \ref{fig51}). 
\\[-1ex]
   \begin{figure}[h!]
    \caption{  \; for\; $p, p'$ and $p''$}
    \noindent\centering
    \includegraphics[width=70mm]{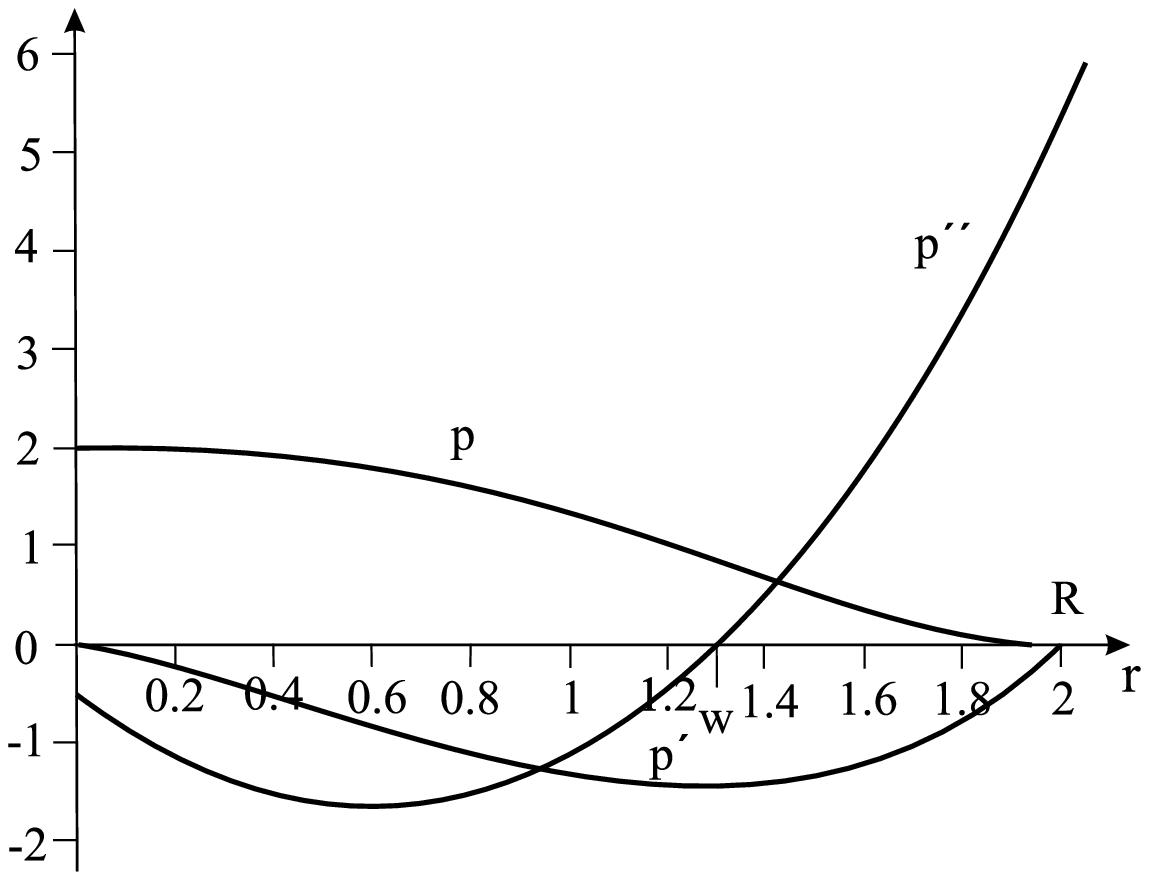}\label{fig51}
    \end{figure}
$~$\\[-1ex]
From \eqref{e5.16} it follows that $p'(r)<0$ for $r\in(0,2)$. This implies that $p\in{\cal D}_R^-(L)$.\\[0ex]
 
In order to calculate $P(r)$, we remark that it is easy to see that  
\begin{equation}\label{e5.18}
L(r^l)=4\pi\left(
-\dfrac{r^{l+2}}{(l+2)(l+3)}+\dfrac{R^{l+2}}{l+2}\right),
\qquad l=0,1,2,\dots.,
\end{equation}
wich directly implies for $l=0,2,3,4$  the bracket terms in the following formular
\\[-4ex]
\begin{multline*}
\hspace{-2ex}P(r)=Lp(r) 
\\=4\pi\biggl[
a_0\left(-\dfrac{r^2}6+\dfrac{R^2}2\right)+
a_2\left(-\dfrac{r^4}{20}+\dfrac{R^4}4\right)+
a_3\left(-\dfrac{r^5}{30}+\dfrac{R^5}5\right)
+a_4\left(-\dfrac{r^6}{42}+\dfrac{R^6}6\right)\biggr] 
 \\=4\pi\left[-\dfrac13r^2+\dfrac{39}{2920}r^4+\dfrac{107}{4380}r^5-\dfrac{15}{2044}r^6+\dfrac{558}{365} \right].
\end{multline*}

Calculating $P'(r)$ and $P''(r)$ and substituting these derivatives into the equality $X(r)=p'(r)P''(r)-p''(r)P'(r)$, we have 
\begin{multline*}
X(r)=\dfrac{4\pi}{746060}\bigl[-1093540r^2+1183812r^3-233688r^4\\ 
-330515r^5+329025r^6-81000r^7\bigr].
\end{multline*}
\\
It is easy to see that the function $X=X(r)$ has the zero at $r_0=0$ and $X(1)<0$, $X(2)>0$. Therefore there is a zero of $X(r)$
 at  $r_1$, $1<r_1<2$ with the approximate value of $r_1\approx w+0.0575585$ (see Diagram \ref{fig52}).
(Occasionally, we work with equivalent fractions of entire numbers to avoid rounding errors of decimal representations.)
\\[-2ex]
 \begin{figure}[h!]
    \caption{  \; for\;  $X, P, P'$ and $P''$}
    \noindent\centering
   \includegraphics[width=70mm]{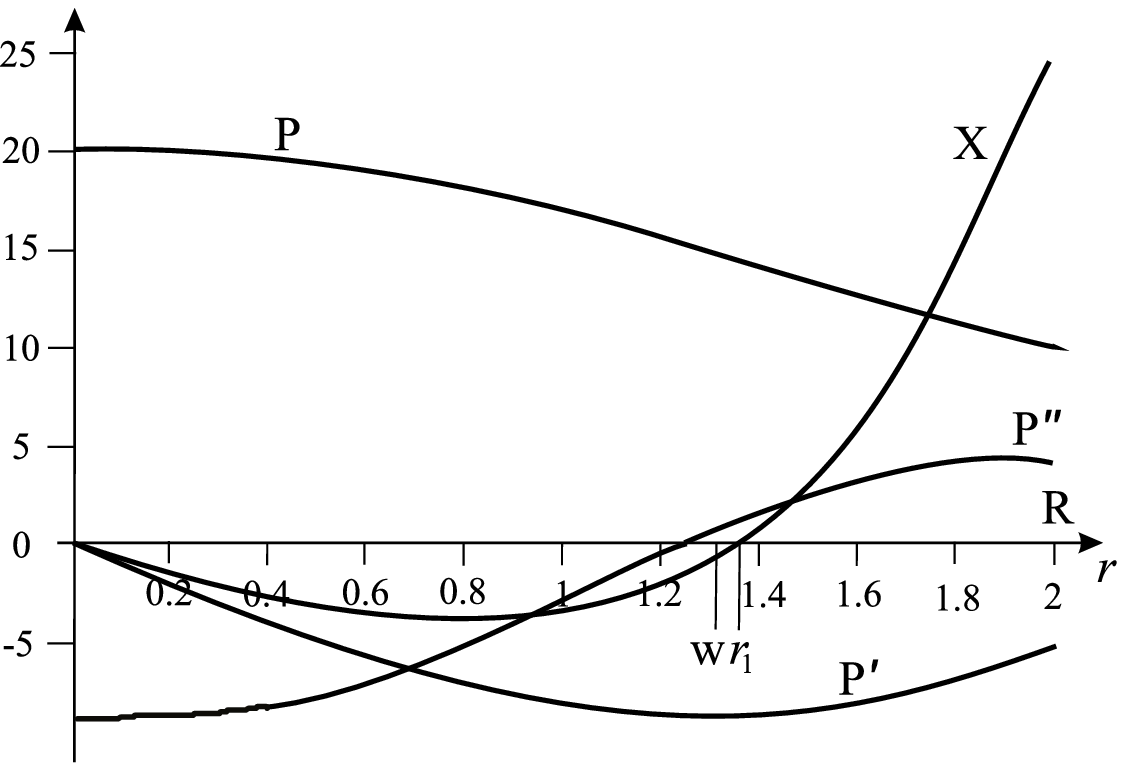}\label{fig52}
    \end{figure}
\\
 Our aim is to show that there exists a point $h\in (0, P(0)-E_0)$ such that $\dfrac{d} {dh}H_{F'_0}(h)<0$. By virtue of 
Theorem \ref{t4.2}\; 2), this means that $p$ is not extendable. Because $p'(r)=0$, \eqref{e4.7} shows that the sign
of $\dfrac{d} {dh}H_{F'_0}(h)$ is equal to the sign of the integral on the right side of \eqref{e4.7}.   
Therefore we have to investigate its integrand
$$ J(r,h):=
\dfrac{X(r)}{\left|P'(r)\right|^2}
\sqrt{\dfrac{P(\Phi(h))-P(R)}{P(\Phi(h))-P(r)} }.
$$
As it was shown in the proof of Theorem \ref{t4.2},\\[-2ex]
\[  \Phi:[0,P(0)-E_0)\to (0,R], h\mapsto \Phi(h)=P^{-1}(h+E_0)  \]
\\[-3ex]    is a $C^2$--diffeomorphysm. 
 If \;$r\in(\Phi(h),R)$, then \;$P(\Phi(h))>P(r)>P(R)$.\\[0ex] 
Hence\; $P(\Phi(h))-P(r)<P(\Phi(h))-P(R)$,\; i.e.\\[-1ex] 
$$
\dfrac{P(\Phi(h))-P(R)}{P(\Phi(h))-P(r)} >1, \quad \text{and} \quad
 \lim\limits_{r\to\Phi(h) }\dfrac{P(\Phi(h))-P(R)}{P(\Phi(h))-P(r)}= \infty,
$$
 \\[-1ex]
and because $J(R)$ and $X(r)$ always have the same sign, that is, \\$J(r)<0$ for $\Phi(h)<r<r_1$\; and\; $J(r)>0$ for $r_1<r<R$.
\\[-0ex]
Since $\Phi$ is a bijection from $[0,P(0)-E_0)$ onto $(0,R]$, for a given $\tilde{r}:=0.01\in(0,R)$, there exists a unique
 $\tilde{h}\in[0,P(0)-E_0)$ such that $\Phi(\tilde{h})=\tilde{r}=0.01$.

By virtue of Lemma \ref{l2.1}, $P(r)\in C^2({\mathbb R}_+)$. Therefore, using Taylor's formula, we have 
$
P(\Phi(\tilde{h}))-P(r)=P'(r+\theta(\Phi(\tilde{h})-r))(\Phi(\tilde{h})-r),
$
where $0<\theta<1$.\ 
We denote
$$
c_0:=\min\limits_{r\in[\tilde{r},R]} |P'(r)|,\quad c_1:=\left|P(\Phi(\tilde{h}))-P(R)\right|,\quad c_2:=\max\limits_{r\in[\tilde{r},R]} |X(r)|.
$$
Clearly, $0<c_i<\infty\, (i=0,1,2)$. Then we have 
$$
J(r,\tilde{h})\le \dfrac{\sqrt{c_1}c_2}{c_0^2}\cdot\dfrac1{\sqrt{\Phi(\tilde{h})-r}}\,.
$$
Therefore the function $J(r,\tilde{h})$ is integrable with respect to $r$ over the interval $(\Phi(\tilde{h}),R)$. 

Now we are going to show that $\dfrac{d} {dh}H_{F'_0}(h)|_{h=\tilde{h}}<0$ by proving the following inequality:\\[-4ex]
\begin{equation}\label{e5.19}
\int\limits_{\Phi(\tilde{h})}^R \, J(r,\tilde{h})\, dr\, <0.
\end{equation}\\[-2ex]
To this end, we introduce two rectangular triangles
$$
T_1:=\{(0.01;0),(0.01;-4),(0.31;0)\}\quad\text{and}\quad 
T_2:=\{(1.3;0),(2;1.21),(2;0)\} $$
with their hypotenuses\\[-4ex]
\begin{align*}
g_1(r)&=\dfrac4{0.3}(r-0.31)\;\text{for}
\; r\in[0.01,0.31]\;\;\text{and}\;\;
g_2(r)=\dfrac{1.21}{0.7}(r-1.3)\;\text{for}\; r\in[1.3, 2]
\end{align*}
\\[-3ex]
(see Diagram \ref{fig53}).\\[2ex]

 \begin{figure}[h!]
    \caption{ \; for  integrand $J$,\; hypotenuses $g_1, g_2$\;  and\; triangles $T_1, T_2$.}
    \noindent\centering
    \includegraphics[width=100mm]{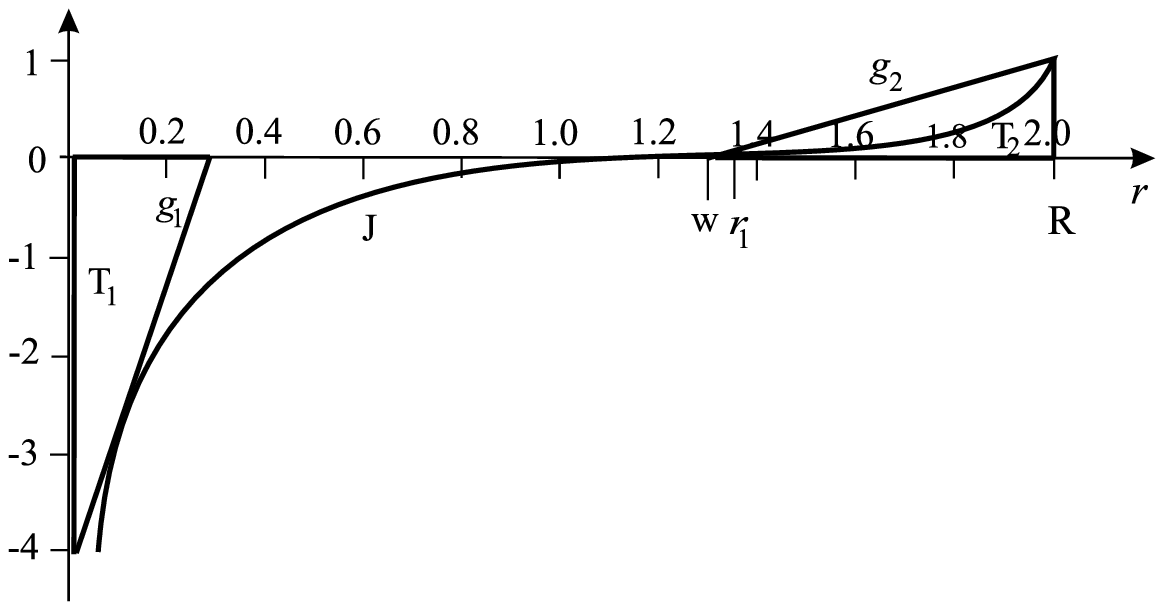}\label{fig53}
    \end{figure}

\medskip
     \renewcommand{\arraystretch}{1.5} 
\vspace{-6mm}
  \begin{table}[h!]
     \caption{ \; for \; $g_1(r)-J(r,\tilde{h})$ } \label{tab5.1}
     \begin{center}
     \begin{tabular}{|c||c|c|c|c|c|c|}
     \hline
\centering $r$ \centering& $0.01$ \centering& $0.1$ \centering& $0.15$ \centering& $0.175$ \centering& $0.2$
\centering& $0.31$ 
    \\
    \hline     
$g_1(r)-J(r,\tilde{h})$ &$\infty$ &$0.832$ &$0.182$ &$0.129$  &$0.173$ &$0.937$
    \\    
  \hline
     \end{tabular}
     \end{center}
     \end{table} 

     \renewcommand{\arraystretch}{1.5} 
%
  \begin{table}[h!]
     \caption{ \;for \;$g_2(r)-J(r,\tilde{h})$} \label{tab5.2}
     \begin{center}
     \begin{tabular}{|c||c|c|c|c|c|}
     \hline
\centering $r$ \centering& $1.3$ \centering& $1.5$ \centering& $1.7$ \centering& $1.9$ \centering& $2$
    \\
    \hline
$g_2(r)-J(r,\tilde{h})$ &$0.027$ &$0.256$ &$0.369$ &$0.225$  &$0.004$ 
    \\    
  \hline
     \end{tabular}
     \end{center}
     \end{table}
$~$\\[-4ex] 
Numerical calculations show (see Charts \ref{tab5.1} and \ref{tab5.2}) that\\[-1ex]
\[ 0>g_1(r)>J(r,\tilde{h})\; \text{for}\;  r\in[0.01,0.31]\quad \text{and}\quad 0< J(r,\tilde{h})<g_2(r) \;\text{for}\; r\in[1.3,2].\]\\[-2ex] 
The area of $T_1$ is $0.60$ and that of $T_2$ is $0.4235$. Therefore, substituting the values\\ 
$\Phi(\tilde{h})=0.01$ and $R=2$, we obtain
\begin{multline*}
\int\limits_{\Phi(\tilde{h})}^R \, J(r,\tilde{h})\, dr\, =
\int\limits_{0.01}^{0.31} \, J(r,\tilde{h})\, dr\, +\,
\int\limits_{0.31}^w \, J(r,\tilde{h})\, dr\, +\,\int\limits_{w}^2 \, J(r,\tilde{h})\, dr\, \\
\le \int\limits_{0.01}^{0.31} \, g_1(r)\, dr\, +\, 0
\, +\,\int\limits_{w}^2 \, g_2(r)\, dr\,=-0.60+0.4235=-0.1765<0.
\end{multline*}
This finishes the proof of the unextendability of $p$.\\

  \begin{figure}[h!]\centering
 \begin{minipage}{100mm}\caption{ for $F_0, F_0', F_0''$ }
  \label{fig54}
    \includegraphics[scale=0.45]{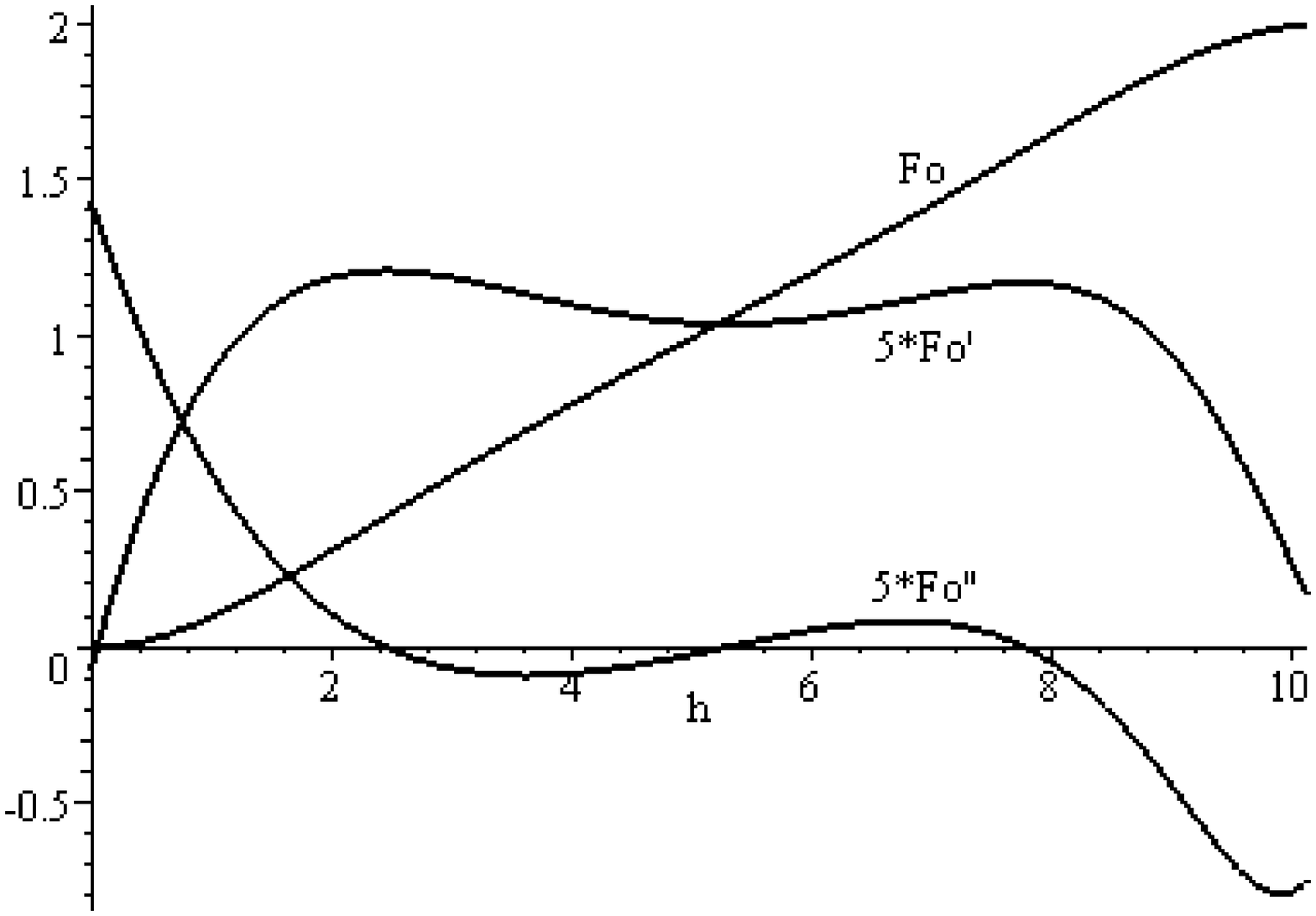}
  \end{minipage}
     \end{figure}
It is interesting to compare for $R=2$\;  $F_0$, $F_0'$, $F_0''$ of Example 5.1 (p extendable) on $h\in[0,\;28/15\pi]$ and
 5.6 (p unextendable) on $h\in[0,\; \approx 10]$.\\
Since in Example 5.1  $F_0' > 0$  and  $F_0" < 0$, then  $F_0 > 0$  is  strictly increasing and  $F_0'$ is  strictly decreasing.
$F_0''$  is strictly decreasing, as shows its formula. \\  
In Example 5.6 it is not possible to calculate $P^{-1}$ explicitly (as in Example\ref{ex5.1}),
 because $P(r)$ is a monotone, nonelemantary polynomial of degree 4. In such a
case there still exists the following possibility: 
\\[1ex]
The set $\{ ( P(r_k), r_{k}) : 0\le r_k\le R; \, k=1,\dots,l ;\, l>0\}$ is contained in the graph of $P^{-1}$,
therefore one can approximate $P^{-1}$ by a
polynomial of degree $m \; (l>2m+1)$ with the method of least squares. 
With this approximation, we can calculate $ F, F_0, F_0', F_0''$ (here we omit the calculations).\\
Diagram \ref{fig54} shows $F_0$ and $F_0', F_0"$ enlarged with the factor $5$, because the value sets of 
$F_0$ and $F_0', F_0"$ are very different.
We see $F_0\ge 0$ and strictly monotone, $F_0'\ge 0$ but not monotone, and  $F_0'' $ is not monotone and changes its sign.
\\[-1ex] 
     \end{lexample}
$~$\\[2ex] 
\section{The direct problem and its equivalent formulation}\label{s6}
We have already mentioned in the introduction in connection with Jeans' theorem, what is
called the direct problem: Given a function $q\in L_{\rm loc}^1[0,T)$, $q>0$ on $[0,T)$
for a sufficiently large interval $[0,T)$ --- under which conditions do there exist
functions $(f_q,\rho,U)$, which form a stationary spherically symmetric $E$-dependent
solution of the (VPS)?

A first answer is given by Lemma \ref{l3.2} \,b): These functions exist if for the
function
    \begin{align*}
F(h)&:=4\pi\sqrt{2}\int_0^{h-E_0}\!\!\!q(s)\sqrt{h-E_0-s}\,ds, \qquad h\in[E_0,P(0)),
    \\
F&\phantom{:}=0 \quad \text{on} \ (-\infty,E_0),
    \end{align*}
the integral equation
    \begin{equation}\label{e6.1}
p(r)=F\circ Lp(r) \quad \text{on} \ \mathbb{R}_{+}
    \end{equation}
has a solution $p\in\mathcal{D}_R^-(L)$. The following Lemma exhibits further properties of
the functions involved.
    \begin{lemma}\label{l6.1}
We may assume $T>P(0)-E_0$ and define
    \begin{equation}\label{e6.2}
F_0(h):=F(h+E_0)=4\pi\sqrt{2}\int_0^h\!\!q(s)\sqrt{h-s}\,ds, \qquad 0\le h<P(0)-E_0.
    \end{equation}

Then $F_0$ is strictly increasing on $[0, P(0)-E_0),$ and so is $F$ on $[E_0, P(0))$. The
solution $p\in\mathcal{D}_R$ of \eqref{e6.1} such that $Lp(R)=E_0$ is in
    \end{lemma}
    \textbf{Proof.}
It follows from Lemma  \ref{lA.4} (b), that $F_0\in AC[0,P(0)-E_0)$ and $F_0$ is strictly
increasing because
    \[
F_0'(h)=\frac{4\pi}{\sqrt{2}}\int_0^h\!\!\frac{q(s)}{\sqrt{h-s}}\, ds>0 \quad \text{a.e.
on} \ [0, P(0)-E_0).
    \]
We show that $p\in\mathcal{D}_R^-(L)$. If $0<r_1<r_2<R$, then $Lp(r_1)>Lp(r_2)$ (Lemma
\ref{l2.2}\, 3)) and $p(r_1)=F\circ Lp(r_1)>F\circ Lp(r_2)=p(r_2)$, so that $p$ is
strictly decreasing on $(0,R)$. Using the relations $F(h)=0$ for $h\in(-\infty, E_0)$,
$Lp(R)=E_0$, and $Lp(r)<Lp(R)$ for $r>R$, we have $p(r)=F\circ Lp(r)=0$
$(r>R)$.\hfill$\square$

We see that the solution of the direct problem is closely connected with solving the
nonlinear integral equation
    \[
p(r)=F\circ Lp(r) \ \text{on} \ \mathbb{R}_{+},
    \]
or --- what is the same after Corollary \ref{co2.1} --- on $(0,R]$. To make this equation
accessible for numerical investigation, we are going to transform it  into a slightly
different form. Because
    \[
F_0\colon[0, P(0)-E_0)\to[0, p(0))
    \]
is a strictly increasing bijection (Lemma \ref{l6.1}), it has a strictly increasing
inverse
    \[
G_0\colon [0,p(0))\to[0,P(0)-E_0)
    \]
and
    \[
F\colon[E_0,P(0))\to[0,p(0))
    \]
has an inverse
    \[
G\colon[0,p(0))\to[E_0,P(0)),
    \]
which has the form $G(h)=G_0(h)+E_0$. Then we get two equivalent statements in the
following Lemma.
    \begin{lemma}\label{l6.2}
Let $p\in\mathcal{D}_R^-(L)$ as in (\ref{e6.1}){\rm,} and $Lp(R)=E_0$. Then
    \[
p=F\circ Lp \Longleftrightarrow Lp-E_0=G_0p \ \text{on} \ [0,R].
    \]
    \end{lemma}
    \textbf{Proof.}
$\Longrightarrow:$ Because $Lp(R)=E_0$, $r\le R$ implies $Lp(r)\ge E_0$. We apply $G$ to 
both parts of the equation $p=F\circ Lp$. Then we get $Gp=Lp$ or $Lp=G_0p+E_0$.

$\Longleftarrow:$ If $Lp-E_0=G_0p$, then $Lp=G_0p+E_0=G(p)$, and application of $F$
yields $F\circ Lp=F\circ Gp=p$.\hfill$\square$

Now let the functions $q$ and $F$ be as in the beginning of this section. Our aim is to
solve the integral equation \eqref{e6.1}, or, what is equivalent according to Lemma
\ref{l6.2}:
    \begin{equation}\label{e6.3}
Lp(r)-Lp(R)=G_0p(r) \ \text{on} \ [0,R].
    \end{equation}
Our fist aim is to consider \eqref{e6.3} at the points $r=R_k$ of the equidistant
partition
    \[
0=R_0<R_1<R_2<\dots<\dots<R_n:=R, \qquad R_k:=k\,\frac{R}{n}\,, \quad k=0,\dots,n, \
\text{of} \ R.
    \]
To this end, we introduce the space of piecewise linear functions
    \[
p_{R_0R_1\dots R_{n-1}R_n}^{x_0x_1\dots x_{n-1}x_n}(r):=-\frac{r-R_k}{R_k-R_{k-1}}\,
x_{k-1}+\frac{r-R_{k-1}}{R_k-R_{k-1}}\cdot x_k \ \text{on}\ [R_{k-1}, R_k],
    \]
$k=1,\dots n$, with the property
    \[
p_{R_0R_1\dots R_{n-1}R_n}^{x_0x_1\dots x_{n-1}x_n} (R_k)=x_k, \qquad k=0,\dots,n.
    \]
A basis for this space is the set
    \[
\left\{p_{R_{k-1}R_kR_{k+1}}^{010}, \qquad k=0,1,\dots,n-1\right\},
    \]
where for the sake of unified notation, we have defined
    \[
p_{R_{k-1}R_kR_{k+1}}^{010}\big|_{k=0}:=p_{R_0R_1}^{10}
    \]
($R_{-1}$ is not defined). Any piecewise linear $p$ with $0=p(R)=p(R_n)=x_n$ can be represented in
the form
    \begin{equation}\label{e6.4}
p(r)=\sum_{k=0}^{n-1}x_k p_{R_{k-1}R_kR_{k+1}}^{010}
    \end{equation}
and its image under $L$ is
    \[
Lp(r)=\sum_{k=0}^{n-1}x_kLp_{R_{k-1}R_kR_{k+1}}^{010}(r).
    \]
If $(x_0,x_1,\dots,x_{n-1})$ is a solution of the system
    \begin{equation}\label{e6.5}    \tag{ANS}
\sum_{k=0}^{n-1}\left(Lp_{R_{k-1}R_kR_{k+1}}^{010}(R_i)-
Lp_{R_{k-1}R_kR_{k+1}}^{010}(R)\right)x_k=G_0(x_i),
    \end{equation}
$i=0,1,\dots,n-1$, and if we define $p$ by \eqref{e6.4} and $Lp(R)=:E_0$, then
    \[
Lp(r)-E_0=G_0\big(p(r)\big) \;\; \text{for} \;  r=R_0,R_1,\dots,R_{n-1}.
    \]
The system is also satisfied for $r=R$ because $Lp(R)=E_0$ and $G_0\big(p(R)\big)=0$. We
call \eqref{eANS} the ``approximating nonlinear system''. It will be the subject of the
next section.
%
%
\section{The approximating nonlinear system (ANS)}\label{s7}
The approximating nonlinear system \eqref{eANS} has the form
    \[
Ax=G_0(x),
    \]
where for a vector $x\in\mathbb{R}^n$, we define
$G_0(x):=\big(G_0(x_i)\big)_{i=0}^{n-1}$ and $A$ is the $n\times{n}$ matrix with coefficients $A_{ik}$,
    \begin{multline}\label{e7.1}
A_{ik}:=Lp_{R_{k-1}R_kR_{k+1}}^{010}(R_i)-Lp_{R_{k-1}R_kR_{k+1}}^{010}(R)=:B_{ik}-C_k,
    \\
i,k=0,1,\dots,n-1.
    \end{multline}
The expressions $Lp_{R_{k-1}R_kR_{k+1}}^{010}(R_i)$ are composed of terms of the form
    \begin{itemize}
\item[(i)] $L_{\rm I}p_{RS}^{01}(r):=Lp_{RS}^{01}(r), \qquad 0\le r\le R<S$,
    \\
\item[(ii)] $L_{\rm I}p_{ST}^{10}(r):=Lp_{ST}^{10}(r), \qquad 0\le r\le S<T$,
    \\
\item[(iii)] $L_{\rm II}p_{RS}^{01}(r):=Lp_{RS}^{01}(r), \hspace*{6.5mm} 0\le R<S\le r$,
     \\
\item[(iv)] $L_{\rm II}p_{ST}^{10}(r):=Lp_{ST}^{10}(r), \hspace*{6.5mm} 0\le S< T\le r$,
    \end{itemize}
where \;  $p_{RS}^{01}(s)=\dfrac{s-R}{S-R} \; \chi_{[R,S]}(s)$, \; 
$p_{ST}^{10}(s)=\dfrac{s-T}{S-T}\,\chi_{[S,T]}(s)$. \;  If $p\in L^\infty(R_+)$ and\\ 
$sp\in L^1(R_+)$, we have
    \[
Lp(0):=\lim_{r\downarrow0}Lp(r)=4\pi\cdot\lim_{r\downarrow0}
\left[\frac1r\int_0^r\!\!p(s)s^2\,ds+\int_r^\infty\!\!p(s)s\,ds\right]=
4\pi\int_0^\infty\!\!p(s)s\,ds.
    \]
For $a,b\in\mathbb{R}$ and $n\ge2$, we shall apply the formula
    \begin{equation}\label{e7.2}
a^n-b^n=(a-b)\big(a^{n-1}+a^{n-2}b+\dots ab^{n-2}+b^{n-1}\big).
    \end{equation}
The calculation of the expressions (i)---(iv) requires the following lemma.

    \begin{lemma}\label{l7.1} $~$\\[1ex]
{\rm(}i{\rm)} Let $0\leqq r\leqq R<S$. Then we have
    \begin{align*}
L_{\rm I}p_{RS}^{01}(r)&=\frac{4\pi}{S-R}\int_R^S\!\!(s-R)s\,ds=\frac{4\pi}{S-R}
\left(\frac{s^3}3-R\,\frac{s^2}2\,\Big|_R^S\right)
    \\
&=\frac{4\pi}{S-R}\left[\frac13\big(S^3-R^3\big)-\frac{R}2\big(S^2-R^2\big)\right]
    \\
&=4\pi\left[\frac13\big(S^2+SR+R^2\big)-\frac{R}2(S+R)\right]\text{\quad with (\ref{e7.2})}
    \\
&=4\pi\left[\frac13\,S^2-\frac16\,SR-\frac16\,R^2\right]=
\frac{4\pi}3\left[S^2-\frac12\,R(S+R)\right].
    \end{align*}
{\rm(}ii{\rm)} Let $0\le r\le S<T$. Then we have
    \begin{align*}
L_{\rm I}p_{ST}^{10}(r)&=4\pi\cdot\frac{(-1)}{T-S}\int_S^T\!\!(s-T)s\,ds
=-\frac{4\pi}{T-S}\left(\frac{s^3}3-T\,\frac{s^2}2\right)\bigg|_S^T
    \\
&=-\frac{4\pi}{T-S}\left[\frac13\,\big(T^3-S^3\big)-\frac12\,T\big(T^2-S^2\big)\right]
    \\
&=-4\pi\left[\frac13\,\big(T^2+TS+S^2\big)-\frac12\,T(T+S)\right]=
\frac{4\pi}3\left[\frac12\,T(T+S)-S^2\right].
    \end{align*}
{\rm(}iii{\rm)} Let $0\le R<S\le r$. Then we have
    \begin{align*}
L_{\rm II}p_{RS}^{01}(r)&=\frac{4\pi}r\,\frac{1}{S-R} \int_R^{S}\!\!(s-R)s^2\,ds=
\frac{4\pi}r\, \frac1{S-R}\left(\frac{s^4}4-R\,\frac{s^3}3\right)\bigg|_R^S
    \\
&=\frac{4\pi}r\,
\frac1{S-R}\left[\frac14\,\big(S^4-R^4\big)-\frac{R}3\,\big(S^3-R^3\big)\right]
    \\
&=\frac{4\pi}r\left[\frac14\,\big(S^3+S^2R+SR^2+R^3\big)-\frac{R}3\,
\big(S^2+RS+R^2\big)\right]
    \\
&=\frac{\pi}r\left[S^3-\frac13\,\big(S^2R+SR^2+R^3\big)\right].
    \end{align*}
{\rm(}iv{\rm)} Let $0\le S<T\le r$. Then we have
    \begin{align*}
L_{\rm II}p_{ST}^{10}(r)&=\frac{-4\pi}r\,\frac1{T-S}\int_S^T\!\!(s-T)s^2\,ds
=-\frac{4\pi}r\, \frac1{T-S}\left(\frac{s^4}4-\frac{Ts^3}3\bigg|_S^T\right)
    \\
&=-\frac{4\pi}r\,\frac1{T-S}\left[\frac14\,\big(T^4-S^4\big)-\frac{T}3\,
\big(T^3-S^3\big)\right]
    \\
&=-\frac{4\pi}r\left[\frac14\,\big(T^3+ST^2+S^2T+S^3\big)-
\frac{T}3\,\big(T^2+ST+S^2\big)\right]
    \\
&=\frac{\pi}r\left[\frac13\,\big(T^3+T^2S+TS^2\big)-S^3\right].
    \end{align*}
    \end{lemma}   
$~$\\[1ex]
    \begin{figure}[h!]
    \caption{}
    \noindent\centering
    \includegraphics[width=70mm]{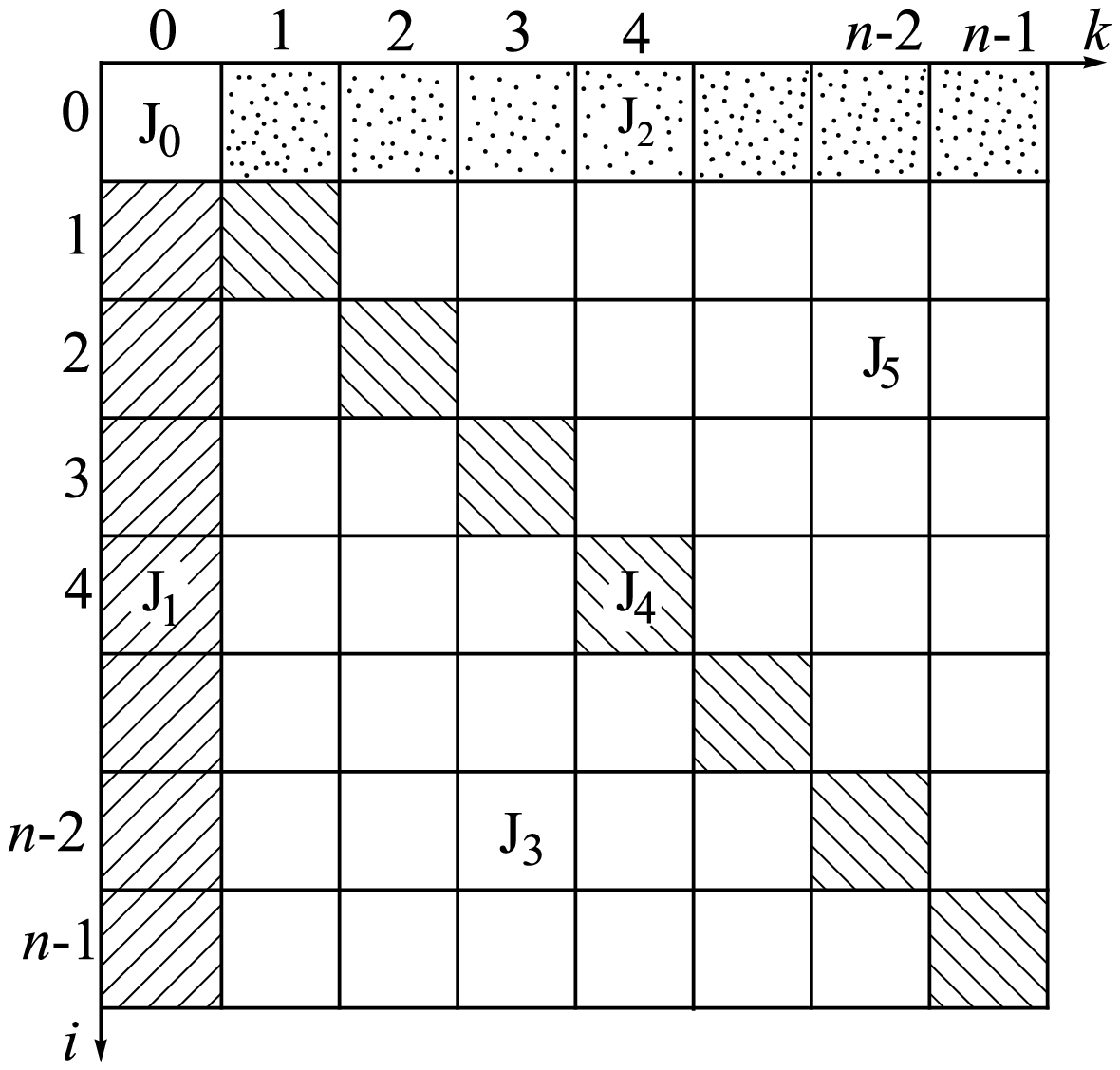}\label{fig71}
    \end{figure}
$~$\\[1ex] 
\noindent In order to calculate the $B_{ik}$ and $C_k$ we divide the square $\{(i,k):
i, k=0,1,2,\dots,n-1\}$ into subsections $J_0,\dots,J_5$:
    \begin{itemize}
\item[] $J_0:=\{(0,0)\}$,
    \\
\item[] $J_1:=\{(i,0)\colon  i=1,\dots,n-1\}$,
    \\
\item[] $J_2:=\{(0,k)\colon  k=1,\dots,n-1\}$,
    \\
\item[] $J_3:=\{(i,k)\colon 2\le i\le n-1$, $1\le k\le i-1\}$,
    \\
\item[] $J_4:=\{(i,k)\colon i=k=1,\dots,n-1\}$,
    \\
\item[] $J_5:=\{(i,k)\colon 1\le i\le k-1, \ 2\le k\le n-1\}$.
    \end{itemize}
    \begin{lemma}\label{l7.2}{\rm(Calculation of the
    $B_{ik}=Lp_{R_{k-1}R_kR_{k+1}}^{010}(R_i)$)}.\\[-1ex]

$J_0=\{0,0\}$. By Lemma {\rm\ref{l7.1},(ii)}, we get
    \[
B_{00}=L_{\rm I}p_{\,0R_1}^{10}(0)=\frac{4\pi}3\,\frac12\,R_1^2
=\frac{2\pi}3\left(\frac{R}n\right)^2.
    \]

$J_1=\{(i,0): i=1,\dots,n-1\}$. By Lemma {\rm\ref{l7.1},(iv)}, we obtain
    \[
B_{i0}=L_{\rm II}p_{\,0R_1}^{10}(R_i)=\frac{\pi}{i\dfrac{R}n}\,
\frac1{3}\left(\frac{R}n\right)^3=\frac{\pi}3\,\frac1i\left(\frac{R}n\right)^2.
    \]

$J_2=\{(0,k): k=1,\dots,n-1\}$. By Lemma {\rm\ref{l7.1},(i),(ii)}, 
we get
    \begin{align*}
B_{0k}&=Lp_{R_{k-1}R_kR_{k+1}}^{010}(0)=L_{\rm I}p_{R_{k-1}R_k}^{01}(0)+L_{\rm
I}p_{R_kR_{k+1}}^{10}(0)
    \\
&=\frac{4\pi}3\left[R_k^2-\frac12\,R_{k-1}\big(R_k+R_{k-1}\big)\right]+
\frac{4\pi}3\left[\frac12\,R_{k+1}\big(R_{k+1}+R_k\big)-R_k^2\right]
    \\
&=\frac{4\pi}3\left(\frac{R}n\right)^2\left[k^2-\frac12\,(k-1)(2k-1)+
\frac12\,(k+1)(2k+1)-k^2\right]
    \\
&=\frac{2\pi}3\left(\frac{R}n\right)^2\left[-\big(2k^2-k-2k+1\big)+
\big(2k^2+k+2k+1\big)\right]=4\pi k\left(\frac{R}n\right)^2.
    \end{align*}

$J_3=\{(i,k): 2\le i\le n-1,$ $1\le k\le i-1\}$. By virtue of Lemma {\rm\ref{l7.1},(iii),(iv)},  we obtain
    \begin{align*}
B_{ik}&=Lp_{R_{k-1}R_kR_{k+1}}^{010}(R_i)=L_{\rm II}p_{R_{k-1}R_k}^{01}(R_i)+ L_{\rm
II}p_{R_kR_{k+1}}^{10}(R_i)
    \\
&=\frac\pi{R_i}\Bigg[R_k^3-\frac13\left(R_k^2R_{k-1}+R_kR_{k-1}^2+ R_{k-1}^3\right)
    \\
&\qquad\qquad+\frac13\left(R_{k+1}^3+R_{k+1}^2R_k+R_{k+1}R_k^2\right)-R_k^3\Bigg]
    \\
&=\frac{\pi}i\,\frac13\left(\frac{R}n\right)^2\big[-\left(k^2(k-1)+
k(k-1)^2+(k-1)^3\right)+(k+1)^3
    \\
&\qquad\qquad+(k+1)^2k+(k+1)k^2\big]
    \\
&=\frac\pi{i}\,\frac13\left(\frac{R}n\right)^2\left[12k^2+2\right]
=4\pi\left(\frac{R}n\right)^2\frac1{i}\,\left(k^2+\frac16\right).
    \end{align*}

$J_4=\{(i,k): i=k=1,\dots,n-1\}$. By virtue of Lemma {\rm\ref{l7.1},(iii),(ii)},  we have
    \begin{align*}
B_{kk}&=Lp_{R_{k-1}R_kR_{k+1}}^{010}(R_k) =L_{\rm II}p_{R_{k-1}R_k}^{01} (R_k)+L_{\rm
I}p_{R_kR_{k+1}}^{10}(R_k)
    \\
&=\frac\pi{R_k}\left[R_k^3-
\frac13\left(R_k^2R_{k-1}+R_kR_{k-1}^2+R_{k-1}^3\right)\right]
    \\
&\qquad\qquad+ \frac{4\pi}3\left[\frac12\,R_{k+1}\big(R_{k+1}+R_k\big)-R_k^2\right]
    \\
&=\frac\pi{k}\left(\frac{R}n\right)^2\left[k^3-
\frac13\left(k^2(k-1)+k(k-1)^2+(k-1)^3\right)\right]
    \\
&\qquad\qquad+
\frac{4\pi}3\left(\frac{R}n\right)^2\left[\frac12(k+1)\left((k+1)+k\right)-k^2\right]
    \\
&=\frac\pi{k}\left(\frac{R}n\right)^2\left[2k^2-\frac43\,k+\frac13\right]
+4\pi\left(\frac{R}n\right)^2\frac13\left[\frac32\,k+\frac12\right]
    \\
&=4\pi\left(\frac{R}n\right)^2\left[k-\frac16+\frac1{12k}\right].
    \end{align*}

$J_5=\{(i,k)\colon 1\le i\le k-1$, $2\le k\le n-1\}$. By virtue of Lemma {\rm\ref{l7.1},(i),(ii)},  we obtain
    \begin{align*}
B_{ik}&=Lp_{R_{k-1}R_kR_{k+1}}^{010}(R_i)=L_{\rm I}p_{R_{k-1}R_k}^{01}(R_i) +L_{\rm
I}p_{R_kR_{k+1}}^{10}(R_i)
    \\
&=\frac{4\pi}3\left[R_k^2-\frac12\,R_{k-1}\big(R_k+R_{k-1}\big)\right]
+\frac{4\pi}3\left[\frac12\big(R_{k+1}\big(R_{k+1}+R_k\big)-R_k^2\right]
    \\
&=\frac{4\pi}3\left(\frac{R}n\right)^2\left[k^2-\frac12(k-1)(2k-1)+
\frac12\big(k^2+2k+1+k^2+k\big)-k^2\right]
    \\
&=\frac{4\pi}3\left(\frac{R}n\right)^2\left[k^2-\frac12\,\big(2k^2-k-2k+1\big)
+\frac12\,\big(2k^2+3k+1\big)-k^2\right]
=4\pi\left(\frac{R}{n}\right)^2k.
    \end{align*}
    \end{lemma}
    \begin{lemma}\label{l7.3}{\rm(Calculation of the
    $C_k:=Lp_{R_{k-1}R_kR_{k+1}}^{010}(R)$)}.\\
From Lemma {\rm\ref{l7.1}, (iv)}, it follows that
    \[
C_0=L_{\rm II}p_{R_0R_1}^{10}(R)=\frac\pi{3R}\cdot\left(\frac{R}n\right)^3.\qquad\qquad\qquad\qquad\quad
    \]
By virtue of Lemma {\rm\ref{l7.1}, (iii),(iv)}, we obtain
    \begin{align*}
C_k&=Lp_{R_{k-1}R_kR_{k+1}}^{010}(R)=L_{\rm II}p_{R_{k-1}R_k}^{01}(R) + L_{\rm
II}p_{R_kR_{k+1}}^{10}(R)
    \\
&=\frac\pi{R}\bigg[R_k^3-\frac13\left(R_k^2R_{k-1}+R_{k-1}^2R_k+R_{k-1}^3\right)
    \\
&\qquad\qquad+\frac13\left(R_{k+1}^3+R_{k+1}^2R_k+R_{k+1}R_k^2\right)-R_k^3\bigg]
    \\
&=\frac\pi{R}\left(\frac{R}n\right)^3\!\bigg[-\frac13\left(k^2(k-1)+k(k-1)^2
\!+\!(k-1)^3\right)
    \\
&\qquad\qquad\quad\;\;\;+\frac13\left((k+1)^3+(k+1)^2k+(k+1)k^2\right)\bigg]
    \\
&=\frac{4\pi}n\left(\frac{R}n\right)^2\left[k^2+\frac16\right].
    \end{align*}
    \end{lemma}
    \begin{corollary}\label{co7.1}
With $A_{ik}=B_{ik}-C_k$ we have
    \[
A_{ik}\big([0,R]\big)=R^2A_{ik}\big([0,1]\big).
    \]
    \end{corollary}
        %
%
%
\section{The numerical analysis of the (ANS)}\label{s8}
The aim of this section is to indicate some of the numerical procedures for solving the system
    \begin{equation}\label{eANS} \tag{ANS}
Ax=G_0(x)
    \end{equation}
with the matrix\; $A=(A_{ik})_{i,k=0}^{n-1}$\,, $A_{ik}:=B_{ik}-C_k$, $i,k=0,\dots,n-1$, 
\\where  $G_0(x):=\big(G_0(x_i)\big)_{i=0}^{n-1}$, \;$x\in (\mathbb{R}_{+})^n$  \; for the scalar function 
\\ $G_0\colon[0,p(0))\to[0,P(0)-E_0)$. The matrix $A$ depends on $n$ $(A=A(n))$ and
can be calculated with the formulas developed in Section \ref{s7}. The solution
$x=(x_0,x_1,\dots,x_{n-1})^T$ represents the values\; $x_0,x_1,\dots,x_{n-1}$ of the
approximation polygon \;$p_{R_0R_1\dots R_{n-1}}^{x_0x_1\dots x_{n-1}}$\; at
$R_0<R_1<\dots<R_{n-1}$ \;with\;\; $p_{R_0R_1\dots R_{n-1}}^{x_0x_1\dots x_{n-1}}(R)=0$\; and\;
$x_0, x_1, \dots, x_{n-1}, x_n=0$,\;\; \\  $R_k:=\dfrac{k}{n}\,R$,\; $k=0,1,\dots,n-1$.

To determine the solution $x$ of the \eqref{eANS} we use Newton's method for the equation
$Ax-G_0(x)=0$. For the convergence of Newton's method a suitable choice of the starting
point $x^{(0)}$ is crucial. In the sequence of partitions
    \[
\pi_{2^i}:=\left\{\frac{k}{2^i}\,R, \ k=0,\dots,2^i\right\}, \qquad i=0,1,2,\dots,
    \]
we use the solution $p_{\pi_{2^i}}$ as starting point for the next partition
$\pi_{2^{i+1}}$ for\\[0.5ex]
$n=2, 4, 8, 16$, $32, 64, 128$, and thus we are able to show that
Kantorovich's criterion \cite[Satz 5.6.3]{10} (which much depends of $x^{(0)}$ and
$\|x^{(0)}-x^{(1)}\|$) yields convergence of the sequences in the following two examples.
For $n=2^0=1$, $Ax-G_0(x)=0$ is a one-dimensional nonlinear equation. To find a solution
of this equation, the method of nested intervals by bisection can be used.\\[2ex]
    %
\\
{\bf Example \ref{ex5.1} of Section \ref{s5}.} $~$ \\[-1ex]
        
In this example we can calculate the operator $G_0$ explicitly. In fact, in
Example~\ref{ex5.1} it was shown that
    \begin{align*}
&F\colon[E_0,P(0)]=\left[\frac8{15}\,\pi R^2, \pi R^2\right]\to[0,1], \ \text{where}
    \\
&F(h)=\sqrt{ah-\frac{20}9}-\frac23\,,\qquad a:=\frac5{\pi R^2}\,.
    \end{align*}
It was also proved that
    \begin{align*}
&F_0\colon\left[0, P(0)-E_0\right]=\left[0,\frac7{15}\pi R^2\right]\to[0,1], \
\text{where}
    \\
&F_0(h)=\sqrt{ah+\frac49}-\frac23\,.
    \end{align*}
Therefore we have
    \begin{align*}
&G_0:=F_0^{-1}\colon [0,1]\to\left[0,\frac7{15}\,\pi R^2\right], \ \text{where}
   \\
&G_0(t)=\frac{\pi R^2}5\left(t^2+\frac43\,t\right).
    \end{align*}
With the described choice of starting points $x^{(0)}$ Newton's method converges for\\
$n=2,4$; for $n=8,16,32,64,128$ Kantorovich's criterion guarantees the convergence. For
each $n$, we have carried out the iterations until the norm of Newton's improvement shows
a relative error with respect to the last iteration of magnitude $10^{-9}$ --- this had
been achieved by less than 5~iterations.

Chart \ref{tab8.1} illustrates the convergence of the polygons $p_n$ as the solutions of the
\eqref{eANS} on the interval $[0,R]$ for $R=8$ towards the strict solution $p$ of the
equation $p(r)=FLp(r)$ or $Lp-E_0=G_0(p)$, which in this case is known as
$p(r)=1-\left(\dfrac{r}{R}\right)^2$. In the center of the chart there are listed the
values $p_n(r_k)$ $(r_k=0,0.5,1,\dots,7.5,8)$ of the approximation polygons for
$n=2,4,\dots,128$. The last column contains the differences $p(r_k)-p_{128}(r_k)$, which
are smaller than $5\cdot 10^{-5}$, and show the pointwise convergence $p_n(r_k)\to
p(r_k)$. The third line from last gives the $L_2$-norms of the differences $p-p_n$,
showing their convergence to $0$ and in particular revealing the fact that doubling the
number of supporting points results in dividing the $L_2$-norm by 4.
The last two lines list the values of the $E_{0n}$ and the relative error
$|E_0-E_{0n}|\big/|E_0|$ in \%. A doubling of supporting point results in a convergence
factor of $1/4$ also here.
%
   \renewcommand{\arraystretch}{1.5} %
    \begin{table}[t]

    \small
    \caption{
    \small  Example \ref{ex5.1} of Section \ref{s5}: \\
 $~$\qquad\qquad\quad\;\; Approximations $p_n(r)$ of $p(r)=1-\left(\dfrac{r}{R}\right)^2$
    for $R=8$ and $n=2$, 4, 8, 16. 32, 64, 128}
    \label{tab8.1}
  \begin{center}
    \begin{tabular}{|*{9}{c|}}
    \hline
$r/n$ &2 &4 &8 &16 &32 &64 &128 &\small$p(r)\!-\!p_{128}(r)$
    \\
    \hline
$0.0$ &$0.83$ &$0.95$ &$0.989$ &$0.99712$ &$0.99928$ &$0.99982$ &$0.99995$&$4.5\!\cdot\!10^{-5}$
    \\
$0.5$ &- &- &- &$0.99322$ &$0.99537$ &$0.99591$ &$0.99605$ &$4.5\!\cdot\!10^{-5}$
    \\
$1.0$ &- &- &$0.973$ &$0.98152$ &$0.98366$ &$0.98420$ &$0.98433$ &$4.5\!\cdot\!10^{-5}$
    \\
$1.5$ &- &- &- &$0.96203$ &$0.06414$ &$0.96467$ &$0.96480$ &$4.4\!\cdot\!10^{-5}$
    \\
$2.0$ &- &$0.89$ &$0.926$ &$0.93474$ &$0.93681$ &$0.93733$ &$0.93746$ &$4.3\!\cdot\!10^{-5}$
    \\
2.5 &- &- &- &0.89965 &0.90167 &0.90217 &0.90230 &$4.2\!\cdot\!10^{-5}$
    \\
3.0 &- &- &0.849 &0.85676 &0.85872 &0.85921 &0.85933 &$4.1\!\cdot\!10^{-5}$
    \\
3.5 &- &- &- &0.80608 &0.80796 &0.80844 &0.80855 &$3.9\!\cdot\!10^{-5}$
    \\
4.0 &0.62&0.71 &0.740 &0.74761 &0.74949 &0.74985 &0.74996 &$3.7\!\cdot\!10^{-5}$
    \\
4.5 &- &- &- &0.68135 &0.68303 &0.68345 &0.68356 &$3.5\!\cdot\!10^{-5}$
    \\
5.0 &- &- &0.601 &0.60730 &0.60886 &0.60925 &0.60934 &$3.2\!\cdot\!10^{-5}$
    \\
5.5 &- &- &- &0.52547 &0.52688 &0.52723 &0.52731 &$2.9\!\cdot\!10^{-5}$
    \\
6.0 &- &0.41 &0.431 &0.43587 &0.43709 &0.43740 &0.43747 &$2.6\!\cdot\!10^{-5}$
    \\
6.5 &- &- &- &0.33850 &0.33951 &0.33976 &0.33982 &$2.1\!\cdot\!10^{-5}$
    \\
7.0 &- &- &0.230 &0.23338 &0.23413 &0.23431 &0.23436 &$1.6\!\cdot\!10^{-5}$
    \\
7.5 &- &- &- &0.12053 &0.12095 &0.12106 &0.12108 &$0.9\!\cdot\!10^{-5}$
    \\
8.0 &0.0 &0.0 &0.0 &0.0 &0.0 &0.0 &0.0 &0.0
    \\
   \hline
\small$\|p\!-\!p_n\|$ &0.45 &0.13 &0.032 &0.00815 &0.00204 &0.00051 &0.00012 &
    \\
    \hline
$E_{0n}$ &79.1&98.9 &105.1 &106.680 &107.094 &107.198 &107.224 &$E_0\!=\!107.233$
    \\
    \hline
Error\%   &26.2&7.73 &2.034 &0.052 &0.013 &0.003 &0.001 &
    \\
    \hline
     \end{tabular}
     \end{center} 
     \end{table}
$~$\\[-4ex]
%
 \begin{figure}[h!]\centering
        \begin{minipage}{60mm}\caption{~to Chart 8.1: Polygon $p_{128}$}
   \label{81}
        \vspace*{8pt}
    \includegraphics[scale=0.565]{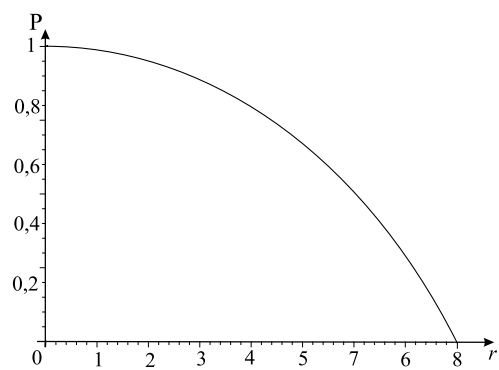}
    %
    \end{minipage}
 \end{figure}
\clearpage
%
$~$\\
{\bf Example \ref{ex5.8} of Section \ref{s5}.} $~$ \\[0ex]
      First we have to compute $G_0(t)=F_0^{-1}(h)$. The equality
$F_0(h)=\dfrac{\pi^2c}{\sqrt{2}}\,h^2$ implies that
$G_0(t)=\dfrac{\sqrt[4]{2}}{\pi\sqrt{c}}\,\sqrt{t}$. To obtain comparable and clearly
representable values $x$ of the solutions  in Chart~\ref{tab8.2} we choose
$c=\sqrt{2}/(4^2 \pi^4 1000)$.

In this example we calculate the solutions $p_n$ as in the first example with Newton's
method. Again Kantorovich's criterion guarantees  its convergence for $n=4,8,\dots,128$.\\[-1.5ex]

The results are shown in Chart \ref{tab8.2}, which differ from 
Chart~\ref{tab8.1} in the following:\\[-3ex]
    \begin{itemize}
\item[1.] The last column shows the relative error (in \%) of\\[-1ex]
$|p_{128}(r_k)-p_{64}(r_k)|\big/ p_{128}(r_k)\cdot100$.
    \\[-3mm]
\item[2.] The third line from last reads
Norm-Error\% $=100\cdot\big|\|p_n\|_2-\|p_{2n}\|_2\big|\big/\|p_{2n}\|_2$.
    \\[-3mm]
\item[3.] The last line reads $E_0-\text{Error \%}=100\cdot|E_{0,n}-E_{0,2n}|\big/
E_{0,2n}$.
    \end{itemize}
$~$\\[-1ex]
%
$~$\\[-8ex]   
  \begin{table}[h!]
    \small
    \caption{\small   Example \ref{ex5.4} of Section \ref{s5}:\\ 
$~$\qquad\qquad\quad\;\;Approximations of $p(r)$ for
    $G_0(t)=4 \pi \sqrt{1000} \sqrt{t}$ for $R=8$ and
     $ n=2$, 4, 8, 16, 32, 64, 128}
    \label{tab8.2}
    \begin{center}
    \begin{tabular}{|*{9}{c|}}
    \hline
$ \hspace{1mm} r / n \hspace{1mm}$ &2 &4 &8 &16 &32 &64 &128 &\large$\substack{|p_{64}\!-\!p_{128}|\\ /p_{128}~\%}$
    \\
    \hline
0.0 &55.55 &71.64 &82.615 &86.30120 &87.30561 &87.56242 &87.62698 &0.074
    \\
0.5 &- &- &- &84.22649 &85.18691 &85.43273 &85.49454 &0.072
    \\
1.0 &- &- &75.227 &78.31755 &79.16157 &79.37742 &79.43169 &0.068
    \\
1.5 &- &- &- &69.44784 &70.12831 &70.30214 &70.34583 &0.062
    \\
2.0 &- &51.37 &56.948 &58.80793 &59.31085 &59.43914 &59.47138 &0.054
    \\
2.5 &- &- &- &47.62660 &47.96641 &48.05298 &48.07473 &0.045
    \\
3.0 &- &- &36.160 &36.93813 &37.14661 &37.19968 &37.21300 &0.036
    \\
3.5 &- &- &- &27.45442 &27.56851 &27.59761 &27.60492 &0.026
    \\
4.0 &18.69 &18.73 &19.345 &19.54416 &19.59812 &19.61189 &19.61535 &0.018
    \\
4.5 &- &- &- &13.29178 &13.31145 &13.31652 &13.31780 &0.010
    \\
5.0 &- &- &8.578 &8.58510 &8.58808 &8.58891 &8.58912 &0.02
    \\
5.5 &- &- &- &5.20257 &5.20247 &5.20170 &5.20152 &0.004
    \\
6.0 &- &3.01&2.917 &2.89867 &2.89447 &2.89345 &2.89319 &0.009
    \\
6.5 &- &- &- &1.41754 &1.41453 &1.41379 &1.41361 &0.013
    \\
7.0 &- &- &0.555 &0.54872 &0.54725 &0.54689 &0.54680 &0.016
    \\
7.5 &- &- &- &0.11996 &0.11958 &0.11949 &0.11947 &0.019
    \\
8.0 &0.0 &0.0 &0.0 &0.0 &0.0 &0.0 &0.0 &0.000
    \\
    \hline
\multicolumn{9}{|l|}{$ \hspace{1mm}\|p_n\|$ \hspace{3mm}80.19 \hspace{3mm}102.73 \hspace{3mm}117.8
\hspace{5mm}122.826 \hspace{5mm}124.204 \hspace{5mm}124.557 \hspace{5mm}124.646
}
    \\[-1ex]
\multicolumn{9}{|l|}{norm-Error\% \hspace{0mm} 21.943 \hspace{3mm}12.77 \hspace{6mm}4.117
\hspace{8mm}1.11\hspace{9mm}0.283 \hspace{8mm}0.071 
}
    \\
    \hline
\multicolumn{9}{|l|}{$E_{0n}$ \hspace{6mm} 2657 \hspace{4mm} 2155 \hspace{3mm}
2081 \hspace{6mm}2065.5 \hspace{5mm} 2061.21 \hspace{5mm}2060.90 \hspace{4mm}2060.75}
    \\[-1ex]
\multicolumn{9}{|l|}{$E_0$-Error\% \hspace{3mm} 23.300 \hspace{3mm}3.562 \hspace{5mm}0.752
\hspace{6mm}0.208 \hspace{9mm}0.015 \hspace{7mm}0.007
}
    \\
    \hline
    \end{tabular}
     \end{center}
     \end{table}
    %
   \begin{figure}[t]\centering
    \begin{minipage}{60mm}\caption{~to Chart 8.2: Polygon $p_{128}$}
    \label{82}
     \vspace*{6pt} 
    \includegraphics[scale=0.65]{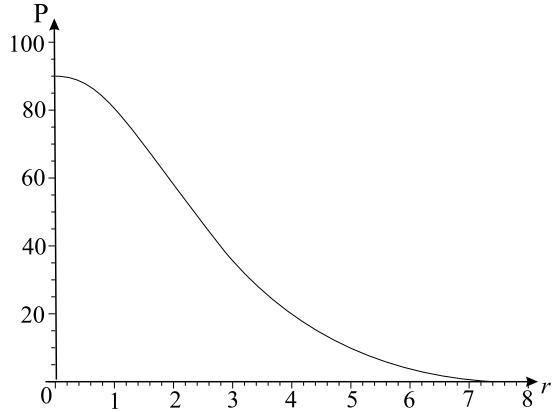}
    %
    \end{minipage}
      \end{figure}


\section{Appendix: Abel's and Eddington's equations}\label{s9}
    \setcounter{section}{1}
    \renewcommand{\thesection}{\Alph{section}}
The goal of this section is the existence proof for Eddington's equation. It is based on
the existence proof for Abel's equation, which we treat first.
\\
We shall use the following notation. For an interval $[0,T)$, we let
    \begin{align*}
L_{\rm loc}^1[0,T)&:=\left\{p\colon[0,T)\to \mathbb{R}; \quad p|_{[0,a]}\in L^1[0,a] \
\text{for all} \ a\in[0,T)\right\},
    \\
AC[0,T)&:=\big\{p\colon[0,T)\to\mathbb{R}; \quad p|_{[0,a]} \ \text{is absolutely
continuous on} \ [0,a]
\\
&\hspace*{41mm}\text{ for all} \ a\in[0,T) \big\}
\; \text{ \cite[ p. 106]{8}}. 
    \end{align*}
We begin with two lemmas and then turn to Abel's equation.
    \begin{lemma}\label{lA.1}
Let $f\in L_{\rm loc}^1[0,T)$. Then the function
    \[
g(x):=\int_0^x\frac{f(s)\,ds}{\sqrt{x-s}}\,ds, \qquad x\in[0,T)
    \]
belongs to $L_{\rm loc}^1[0,T)$.
    \end{lemma}
    \textbf{Proof.}
We may assume that $f\ge0$. Using the equality
    \[
\int_a^b\!\!\frac{d\sigma}{\sqrt{b-\sigma}\,\sqrt{\sigma-a}}=\pi, \qquad 0\leq a<b,
    \]
and Fubini's Theorem, we obtain
    \begin{align*}
\infty>\int_0^x\!\!f(s)\,ds&=\frac1\pi\int_0^x\!\!f(s)\int_s^x\!\!
\frac1{\sqrt{x-\sigma}\,\sqrt{\sigma-s}}\,d\sigma\,ds
    \\
&= \frac1\pi\int_0^x \left(\int_0^\sigma \frac{f(s)}{\sqrt{\sigma-s}}\,ds\right)
\frac1{\sqrt{x-\sigma}}\,d\sigma
    \\[1mm]
&=\frac1\pi\int_0^x\!\!g(\sigma)\frac1{\sqrt{x-\sigma}}\,d\sigma.
    \end{align*}
It follows $\left(\sigma\to g(\sigma)\dfrac1{\sqrt{x-\sigma}}\right)\in L_{\rm
loc}^1[0,T)$ and $g\in L_{\rm loc}^1[0,T)$.\hfill$\square$
    \begin{lemma}\label{lA.2}
Let $f,g\in L_{\rm loc}^1[0,T)$. Then we have
    \begin{equation}\label{eA.1}
\int_0^x\!\!\frac{f(s)}{\sqrt{x-s}}\,ds=g(x) \quad \text{a.e. on} \ (0,T)
    \end{equation}
if and only if
    \begin{equation}\label{eA.2}
\int_0^x\!\!f(s)\,ds=\frac1\pi\int_0^x\!\!\frac{g(s)}{\sqrt{x-s}}\,ds \quad \text{on} \
(0,T).
    \end{equation}
    \end{lemma}
    \textbf{Proof.}
Let \eqref{eA.1} hold. Then
    \begin{align*}
\int_0^x\!\!\frac{g(s)}{\sqrt{x-s}}\,ds&=\int_0^x\left(
\int_0^s\!\!\frac{f(\sigma)}{\sqrt{s-\sigma}}\,d\sigma\right)\frac1{\sqrt{x-s}}\,ds
    \\[1mm]
&=\int_0^x\!\!f(\sigma)\left(\int_\sigma^x\!\!\frac1{\sqrt{x-s}\,
\sqrt{s-\sigma}}\,ds\right)\,d\sigma=\pi\int_0^x\!\!f(\sigma)d\sigma,
    \end{align*}
i.e., \eqref{eA.2} is fulfilled.
\\
Now assume \eqref{eA.2}.
Let $h(x):=\displaystyle\int_0^x\dfrac{f(s)}{\sqrt{x-s}}\,ds-g(x)$. We need to show
$h=0$. Lemma \ref{lA.1} gives $h\in L_{\rm  loc}^1[0,T)$. As in the first part of
the proof, using \eqref{eA.2}, we have
    \begin{align*}
\int_0^x\!\!\frac{h(s)}{\sqrt{x-s}}\,ds&=\int_0^x\!\!\frac1{\sqrt{x-s}}
\int_0^s\!\!\frac{f(\sigma)}{\sqrt{s-\sigma}}\,d\sigma\,ds-\int_0^x\!\!
\frac{g(s)}{\sqrt{x-s}}\,ds
    \\[1mm]
&=\int_0^x\left(\int_\sigma^x\frac{ds}{\sqrt{x-\sigma}\,\sqrt{\sigma-s}}\right)
f(\sigma)\,d\sigma-\int_0^x\!\!\frac{g(s)}{\sqrt{x-s}}\,ds=
    \\[1mm]
&=\pi\int_0^x\!\!f(\sigma)\,d\sigma-\int_0^x\!\!\frac{g(s)}{\sqrt{x-s}}\,ds=0.
    \end{align*}
Hence
    \begin{align*}
\pi\int_0^x\!\! h(s)\,ds&=\int_0^x\!\! h(s)\int_s^x\!\!
\frac{1}{\sqrt{x-\sigma}\,\sqrt{\sigma-s}}\,d\sigma\,ds=
    \\[1mm]
&=\int_0^x\!\!\frac1{\sqrt{x-\sigma}}\left(\int_0^\sigma\!\!
\frac{h(s)}{\sqrt{\sigma-s}}\, ds\right)d\sigma=0, \qquad x\in[0,T),
    \end{align*}
and $h=0$ follows.\hfill$\square$ \\[-1ex]

We now consider the solvability of Abel's equation.
    \begin{lemma}{\rm (Existence and uniqueness for Abel's equation).}\label{lA.3}
\\[1ex]
\noindent
{\rm(}a{\rm)} Let $g\in L_{\rm loc}^1[0,T)$ and assume

(i) $G\in AC[0,T)${\rm,} where
$\displaystyle G(x):=\int_0^x\!\!\dfrac{g(s)}{\sqrt{x-s}}\,ds$, $~$ \\

(ii) $G(0)=0$.\\

Then $f$ defined by
    \[
f(x):=\frac1\pi \,G'(x)
    \]
\qquad is the unique solution of Abel's equation
    \begin{equation}\label{eA.3}
g(x)=\int_0^x\!\!\frac{f(s)}{\sqrt{x-s}}\,ds, \qquad x\in[0,T).
    \end{equation}
    \noindent
 {\rm(}b{\rm)} Conversely{\rm,} if $f\in L_{\rm loc}^1[0,T)$ and $g$ satisfies
\eqref{eA.3}{\rm,} then $g\in AC[0,T)${\rm,} {\rm (}i{\rm),} {\rm(}ii{\rm)}
hold{\rm,} \;\; and
    \[
f(x)=\frac1\pi\,G'(x), \qquad x\in[0,T).
    \]
    \end{lemma}
    \textbf{Proof.}
(a) By assumption, $f\in L_{\rm loc}^1[0,T)$, and we have
    \[
\int_0^x\!\!f(s)\,ds=\frac1\pi\int_0^x\!\!G'(s)\,ds=\frac1\pi\big(G(x)-0\big)=
\frac1\pi\int_0^x\frac{g(s)}{\sqrt{x-s}}\, ds, \qquad x\in(0,T).
    \]
Lemma \ref{lA.2} then shows that $f$ solves \eqref{eA.3}. The uniqueness follows  from
(b).

(b) It follows from Lemma \ref{lA.1} that $g\in L_{\rm loc}^1[0,T)$. We define
    \[
G(x):=\int_0^x\!\!\frac{g(s)}{\sqrt{x-s}}\,ds.
    \]
Since $f$ satisfies \eqref{eA.3}, Lemma \ref{lA.2} gives
    \[
\int_0^x\!\!f(s)\,ds=\frac1\pi\, G(x), \qquad x\in[0,T).
    \]
Hence $G\in AC[0,T)$, $G(0)=0$, and $f(x)=\dfrac1\pi\,G'(x)$, $x\in(0,T)$, which proves
the uniqueness assertion in (a).\hfill$\square$

    \begin{lremark}\label{rA.1}
The example $g(s)=\dfrac1{\sqrt{s}}$\,, $\displaystyle
G(x)=\int_0^x\!\!\dfrac{ds}{\sqrt{s}\,\sqrt{x-s}}=\pi$ shows that (ii) does not
necessarily follow from (i).

Now we treat the solvability of Eddington's equation.
    \end{lremark}
    \begin{lemma}{\rm (Existence and uniqueness for Eddington's equation).} \label{lA.4}
\\[1ex]
{\rm(}a{\rm)} Let $g\in AC[0,T)$, $g(0)=0$, and assume that
    \begin{itemize}
\item[\rm (i)] $H_{g'}\in AC[0,T)${\rm,} where
$\displaystyle H_{g'}(x):=\int_0^x\!\!\dfrac{g'(s)}{\sqrt{x-s}}\,ds$,
    \\[-1ex]
\item[\rm (ii)] $H_{g'}(0)=0$.
    \end{itemize}
Then $f$ defined by
    \begin{equation}\label{eA.4}
f(x):=\frac2\pi\,H_{g'}'(x)
    \end{equation}
is the unique solution of Eddington's equation
    \begin{equation}\label{eA.5}
g(x)=\int_0^x\!\!f(s)\sqrt{x-s}\, ds.
    \end{equation}
{\rm(}b{\rm)} Conversely{\rm,} if $f\in L_{\rm loc}^1[0,T)${\rm,} and $g$ satisfies
\eqref{eA.5}, then $g\in AC[0,T)${\rm,} $g(0)=0$, {\rm(i),  (ii)}  hold, 
and $f$ has the representation \eqref{eA.4}. In addition, $\displaystyle
g'(x)=\frac12\int_0^x\frac{f(s)}{\sqrt{x-s}}\, ds$.
    \end{lemma}
    \textbf{Proof.}
(a) Equation \eqref{eA.4}, partial integration, (i) (ii), and
Fubini's theorem yield
    \begin{align*}
\int_0^x\!\!f(s)\sqrt{x-s}\, ds&=\frac2\pi\int_0^x\!\! H_{g'}'(s)\sqrt{x-s}\,ds=
\frac1\pi\int_0^x\!\!\frac{H_{g'}(s)}{\sqrt{x-s}}\,ds
    \\[1mm]
&=\frac1\pi\int_0^x\!\!\frac1{\sqrt{x-s}}\left(\int_0^s\frac{g'(\sigma)}{\sqrt{s-\sigma}}
\, d\sigma\right)ds
    \\[1mm]
&=\frac1\pi\int_0^x\left(\int_\sigma^x\!\!
\frac{ds}{\sqrt{x-s}\,\sqrt{s-\sigma}}\right)g'(\sigma)d\sigma=g(x),
    \end{align*}
that is, \eqref{eA.5}. Uniqueness follows from (b).

(b) Let $f\in L_{\rm loc}^1[0,T)$, and $g$ satisfy \eqref{eA.5}. We show that $g\in
AC[0,T)$ and $g(0)=0$. We formally define
    \[
h(x):=\int_0^x\!\!\frac{f(s)}{\sqrt{x-s}}\,ds.
    \]
By Lemma \ref{lA.1}, $h$ exists a.e. on $[0,T)$ and $h\in L_{\rm loc}^1[0,T)$. In fact,
equality \eqref{eA.5} and Fubini's theorem imply
    \begin{align*}
2g(x)&=2\int_0^x\!\!f(\sigma)\sqrt{x-\sigma}\,d\sigma=
\int_0^x\!\!f(\sigma)\left(\int_\sigma^x\!\!\frac{ds}{\sqrt{s-\sigma}}\right)d\sigma
    \\
&=\int_0^x\left(\int_0^s\!\!\frac{f(\sigma)}{\sqrt{s-\sigma}}\, d\sigma\right)ds=
\int_0^x\!\! h(s)\,ds.
    \end{align*}
Hence $g\in AC[0,T)$, and $g(0)=0$. Furthermore,
    \[
2g'(x)=h(x)=\int_0^x\!\!\frac{f(s)}{\sqrt{x-s}}\,ds,
    \]
and from Lemma \ref{lA.4} it follows that
    \[
\int_0^x\!\!f(s)\,ds=\frac2\pi\int_0^x\!\!\frac{g'(s)}{\sqrt{x-s}}\,ds=\frac2\pi\,
H_{g'}(x).
    \]
Hence $H_{g'}\in AC[0,T)$, and $H_{g'}(0)=0$. Differentiating the last equality, we get
\eqref{eA.4}, which also shows the uniqueness in (a).
   \hfill$\square$
   %

  \renewcommand{\thesection}{{10}}
    \section{Suggestions for further work}\label{s10}
    In Example \ref{ex5.9} it was constructed a function $p$ satisfying the assumptions of Theorem \ref{t4.2} such that according
 to numerical calculations the function $q$ can have negative values, i.e. $p$ is nonextendable. It would be interesting to give a 
rigorous proof of this result.  
\\A further question is the extension of the present work to the case of cylindrical symmetry.
%

\section*{Acknowledgements} 
The work of the third autor was supported by the Ministry of Science and Education of 
Russian Federation, project number FSSF--2020--0018. 
\\ [1ex]
The autors are grateful to Academician of RAS  Prof. V.V. Kozlow for recommending the publication of a preview of 
the present work in {\it Doklady Mathematics} 
(see \cite{2a}) .

    %
%
%

$~$\\[-4ex]

\end{document}